\documentclass[]{emulateapj}
\usepackage{graphicx}

\begin{document}

\title{New evolutionary sequences for  hot H-deficient white dwarfs on
       the basis of a full account of progenitor evolution}

\author{L. G. Althaus$^{1,4}$,
        J. A. Panei$^{1,2}$,
        M. M. Miller Bertolami$^{1,2,3}$,
        E. Garc\'{\i}a--Berro$^{4,5}$,
        A. H. C\'orsico$^{1,2}$,\\
        A. D. Romero$^{1,2}$,
        S. O. Kepler$^{6}$,
        and 
        R. D. Rohrmann$^{7}$}

\affil{$^1$Facultad de Ciencias Astron\'omicas y Geof\'{\i}sicas, 
           Universidad Nacional de La Plata, 
           Paseo del Bosque s/n, 
           (1900) La Plata, 
           Argentina\\
       $^2$Instituto de Astrof\'{\i}sica de La Plata, 
           IALP (CCT La Plata), 
           CONICET-UNLP\\
       $^3$Max Planck Institut f\"ur Astrophysik,
           Karl-Schwarzschild-Str. 1, 
           85748 Garching, 
           Germany\\
       $^4$Departament de F\'\i sica Aplicada,
           Universitat Polit\`ecnica de Catalunya,
           c/Esteve Terrades 5, 
           08860 Castelldefels,
           Spain\\
       $^5$Institute for Space Studies of Catalonia, 
           c/Gran Capit\`a 2--4, 
           Edif. Nexus 104, 
           08034 Barcelona,
           Spain\\
       $^6$Instituto de F\'\i sica, 
           Universidade Federal do Rio Grande do Sul, 
           91501-970 Porto Alegre, RS, 
           Brazil\\
       $^7$Observatorio Astron\'omico, 
           Universidad Nacional de C\'ordoba, 
           Laprida 854, 
           (5000) C\'ordoba, 
           Argentina}
 
\email{althaus@fcaglp.unlp.edu.ar}

\begin{abstract} 
We present full evolutionary calculations appropriate for the study of
hot hydrogen-deficent  DO white  dwarfs, PG 1159  stars, and  DB white
dwarfs.   White dwarf  sequences  are  computed for  a  wide range  of
stellar  masses  and helium  envelopes  on  the  basis of  a  complete
treatment of the evolutionary  history of progenitors stars, including
the core hydrogen and helium burning phases, the thermally-pulsing AGB
phase, and the born-again episode that is responsible for the hydrogen
deficiency.   We  also  provide  colors  and magnitudes  for  the  new
sequences for  $T_{\rm eff} < 40\,000$  K, where the  NLTE effects are
not  dominant. These new  calculations provide  an homogeneous  set of
evolutionary tracks  appropriate for  mass and age  determinations for
both PG 1159 stars and DO white dwarfs.  The calculations are extended
down to  an effective temperature of  7\,000 K.  We  applied these new
tracks to  redetermine stellar masses and  ages of all  known DO white
dwarfs  with spectroscopically-determined  effective  temperatures and
gravities, and compare them with previous results. We also compare for
the first time consistent mass  determinations for both DO and PG 1159
stars,  and find  a considerably  higher mean  mass for  the  DO white
dwarfs.   We discuss  as well  the  chemical profile  expected in  the
envelope of  variable DB  white dwarfs from  the consideration  of the
evolutionary  history  of   progenitor  stars.   Finally,  we  present
tentative evidence  for a  different evolutionary channel,  other than
that  involving  the  PG  1159   stars,  for  the  formation  of  hot,
hydrogen-deficient white dwarfs.
\end{abstract}         



\section{Introduction}
\label{intro}

White dwarf  stars are  the end-product of  the evolution of  the vast
majority of stars.  Indeed, more than  97\% of all stars in our Galaxy
are expected to end their lives as white dwarfs.  For this reason, the
present population of white dwarfs contains valuable information about
the evolution of individual stars, the previous history of the Galatic
populations (Isern et al.  1998; Torres  et al.  2002) and the rate of
star  formation (D\'\i az--Pinto  et al.   1994).  In  addition, white
dwarfs have potential applications  as reliable cosmic clocks to infer
the age of a wide variety of stellar populations (Winget et al.  1987;
Garc\'\i  a--Berro et al.   1988; Hansen  et al.  2007), and  to place
constraints on  elementary particles (C\'orsico et al.  2001; Isern et
al. 2008).  Recent reviews  on the properties  and evolution  of white
dwarfs and of their applications  are those of Winget \& Kepler (2008)
and Fontaine \& Brassard (2008).

Traditionally,  white dwarfs  have been  classified into  two distinct
families  according to the  main constituent  of their  surface: those
with a hydrogen (H)-dominated atmosphere  --- the DA spectral type ---
and those with  a helium (He)-rich surface composition  --- the non-DA
white dwarfs.   The latter, that  comprise about $\sim 20$\%  of known
white  dwarfs,  are usually  divided  into  three  subclasses: the  DO
spectral type (with effective temperatures 45\,000 K $\leq T_{\rm eff}
\leq 200\,000$ K) that shows relatively strong lines of singly ionized
He  (He{\sc  ii}), the  DB  type (11\,000  K  $\leq  T_{\rm eff}  \leq
30\,000$ K), with strong neutral He (He{\sc i}) lines, and the DC, DQ,
and DZ types ($T_{\rm eff} <  11\,000$ K) showing traces of carbon and
metals in their spectra.  As a  DO white dwarf evolves, the He{\sc ii}
recombines to form He{\sc i},  ultimately transforming into a DB white
dwarf.  The transition  from DO to the cooler  DB stage is interrupted
by the non-DA gap (that occurs at 30\,000 K $ < T_{\rm eff} < 45\,000$
K) where  few objects with H-deficient atmospheres  have been observed
(Eisenstein et al. 2006).  To this  list, we have to add the discovery
of a new white  dwarf spectral type with carbon-dominated atmospheres,
the ``hot DQ'' white dwarfs,  with $T_{\rm eff}\sim 20\,000$ K (Dufour
et al.  2007, 2008).

DO white dwarfs are usually thought to be the progeny of PG 1159 stars
(Dreizler \& Werner 1996; Unglaub  \& Bues 2000; Althaus et al. 2005),
hot  stars with  H-deficient and  He-,  C- and  O-rich surface  layers
(Werner  \&  Herwig 2006).   A  large fraction  of  PG  1159 stars  is
believed to be  the result of a born-again episode,  i.e., a very late
thermal pulse  (VLTP) experienced  by a white  dwarf during  its early
cooling phase  (Fujimoto 1977; Sch\"onberner 1979; Iben  et al.  1983;
Althaus et al.  2005).  During the  VLTP, most of the H content of the
remnant is violently burned (Herwig  et al.  1999; Miller Bertolami et
al. 2006). As  a result, the remnant is forced  to evolve rapidly back
to the AGB and finally into  the central star of a planetary nebula as
a  hot H-deficient object  and with  a surface  composition resembling
that of the intershell region chemistry of AGB star models --- typical
mass abundances are $X_{\rm He}\simeq 0.33$, $X_{\rm C}\simeq 0.5$ and
$X_{\rm O}\simeq 0.17$, though  notable variations are found from star
to  star   (Dreizler  \&   Heber  1998;  Werner   2001).   Eventually,
gravitational settling  acting during the early stages  of white dwarf
evolution causes He to float and heavier elements to sink down, giving
rise to an  He-dominated surface, and turning the PG  1159 star into a
DO  white  dwarf (Unglaub  \&  Bues  2000).   During this  stage,  the
evolution of the  star is dictated essentially by  neutrino losses and
the release  of gravothermal energy (O'Brien \&  Kawaler 2000; Althaus
et al.  2005).  Because these white  dwarfs are the  hottest ones, the
evolution  during  the  DO  stage  proceeds very  fast,  with  typical
evolutionary time scales $\sim$1 Myr.

Several  works have been  devoted to  the identification  and spectral
analysis of H-deficient  objects like PG 1159 and  DO white dwarfs ---
see Dreizler \& Werner (1996),  Dreizler et al.  (1997) and references
therein for  earlier studies  of spectroscopically confirmed  DO white
dwarfs.  Particularly  relevant is the star  KPD0005+5106, the hottest
known DO white dwarf with an effective temperature of about 200\,000 K
(Werner et al.  2008a),  much higher than previously thought (Dreizler
\& Werner  1996).  With  the advent of  large surveys, like  the Sloan
Digital   Sky   Survey   (York   et   al.    2000)   the   number   of
spectroscopically-identified    H-deficient   stars    has   increased
considerably.  For  instance, Krzesi\'nski  et al. (2004)  reported 15
spectroscopically-identified  DO  stars  ---  13  of  which  were  new
discoveries --- from the SDSS Data Release 1 (Abazajian et al.  2003),
which contains  $\approx$2\,500 white dwarfs (Kleinman  et al.  2004).
Also, Eisenstein et  al. (2006) reported 31 DO white  dwarfs and 10 PG
1159 stars, from their catalog  based on SDSS DR4 (Adelman-McCarthy et
al.  2006).   H\"ugelmeyer et al.  (2005)  performed spectral analyses
of 16 hot  H-deficient stars from DR1, DR2 and DR3  catalogs , where 9
were  classified  as  DO and  5  as  PG  1159 stars.   More  recently,
H\"ugelmeyer et al.  (2006) extended this work to 13 DO  and 4 PG 1159
stars  --- see  also Werner  et  al.  (2004).   These authors  derived
effective temperatures, surface gravities and stellar masses for their
sample, and for that of H\"ugelmeyer et al.  (2005), and discussed the
evolutionary connection between DO white dwarfs and PG 1159 stars.

To infer  the mass and age  of observed DO  white dwarfs, evolutionary
calculations  for these  stars  are required.   However,  most of  the
existing  calculations  that  are  employed in  the  determination  of
evolutionary  parameters  of  non-DA  white dwarf  stars  (Wood  1995;
Benvenuto  \&  Althaus 1999)  are  not  entirely  consistent with  the
evolutionary history that leads to the formation of H-deficient stars.
Even though  significant deviations are not expected  at low effective
temperatures, this  lack of consistency becomes relevant  for the high
effective  temperatures characteristic  of  hot DO  white dwarfs.   In
addition,  the lack  of evolutionary  calculations covering  the whole
stage from the domain of luminous PG 1159 to the hot white dwarf stage
prevents from a consistent mass  determination for the DOs and PG 1159
stars (Dreizler \& Werner 1996).  To our knowledge, the only exception
is the study of Althaus et  al. (2005), who computed the formation and
evolution of  a $0.588 \,M_{\sun}$  white dwarf remnant from  the ZAMS
through the born-again  and PG 1159 stages to  the white dwarf regime.
This paper  is precisely intended to  fill this gap,  by computing new
evolutionary  tracks and  mass-radius relations  appropriate  for both
non-DA white  dwarfs and PG 1159  stars. Sequences are  derived from a
full  and  self-consistent  treatment  of  the  complete  evolutionary
history of progenitor stars  with different masses evolved through the
born-again  episode.   In  this  way,  the  white  dwarf  evolutionary
calculations  presented  here  are  not  affected  by  inconsistencies
arising from  artificial procedures  to generate starting  white dwarf
configurations.   In  addition, the  computation  of the  evolutionary
history of  progenitor stars  gives us  the amount of  He left  in the
white dwarf as well as the chemical profiles expected not only for the
carbon-oxygen  core, but  also  for the  partially degenerate  regions
above the core, of relevance for the white dwarf cooling phase.

The  calculations  presented here  constitute  an  homogeneous set  of
evolutionary tracks  aimed to  study both PG  1159 and DO  white dwarf
stars, as  well as  the DB  white dwarfs.  The  paper is  organized as
follows. In  Sect.  \ref{Computational} we describe  the main physical
inputs to  the models.  We  also describe the main  characteristics of
the initial models and  some details of the evolutionary computations.
In  Sect.  \ref{results}  we present  the results  for DO  white dwarf
evolutionary sequences. We apply  the sequences to redetermine stellar
masses and ages of observed DO white dwarfs, and we derive mass-radius
relations.  In addition, we show the  results for the case of DO white
dwarf sequences with very thin  He-rich envelopes.  We close the paper
in  Sect.   \ref{conclusiones}   by  discussing  and  summarizing  the
implications of our results,  particularly regarding the origin of the
observed DO white dwarfs.

\begin{table}
\centering
\caption{Initial and final stellar mass (in solar units), and the total 
         mass  of  He  content  left   in  the  white  dwarf  for  the
         evolutionary sequences considered in this work.}
\begin{tabular}{llc}
\hline
\hline
$M_{\rm WD}$ &$M_{\rm ZAMS}$  &  $M_{\rm He}$ \\
\hline
 0.515   & 1.00 & 0.0219 \\
 0.530   & 1.00 & 0.0090 \\
 0.542   & 1.00 & 0.0072 \\
 0.565   & 2.20 & 0.0067 \\
 0.584   & 2.50 & 0.0060 \\
 0.609   & 3.05 & 0.0058 \\
 0.664   & 3.50 & 0.0036 \\
 0.741   & 3.75 & 0.0019 \\
 0.870   & 5.50 & 0.0009 \\
\hline
\hline
\end{tabular}
\label{tableini}
\end{table}  


\section{Computational details}
\label{Computational}

\subsection{Input physics}

The evolutionary  calculations presented in  this work were  done with
the {\tt LPCODE} stellar evolutionary code we employed in our previous
studies  on the  formation of  the H-deficient  stars through  late He
flashes, like PG 1159 and  extreme horizontal branch stars (Althaus et
al. 2005;  Miller Bertolami \&  Althaus 2006; Miller Bertolami  et al.
2008).  More recently,  the code was used to  compute the formation of
hot DQ  white dwarfs (Althaus et  al. 2009b) as well  as the formation
and  evolution   of  He-core   white  dwarfs  with   high  metallicity
progenitors (Althaus  et al. 2009c).   Details of {\tt LPCODE}  can be
found in these works. In particular, the code considers a simultaneous
treatment of  non-instantaneous mixing and burning  of elements, which
is  of primary importance  for the  calculation of  chemical abundance
changes during  the short-lived evolutionary  stages characteristic of
unstable burning  episodes, like the born-again stage,  from which our
starting  H-deficient  white  dwarf  configurations are  derived,  see
2.1. Nuclear  reaction rates  are from Caughlan  \& Fowler  (1988) and
Angulo et al.   (1999), and are detailed in Althaus  et al. (2005). In
particular,  the  $^{12}$C($\alpha,\gamma)^{16}$O  reaction  rate  was
taken from Angulo et al. (1999), which is about twice as large as that
of Caughlan  \& Fowler (1988).  A moderate, diffusive  overshooting in
the  core  and in  the  envelope  is  allowed during  pre-white  dwarf
evolution.

For the white dwarf regime,  we considered the following main physical
ingredients.    Neutrino   emission   rates   for  pair,   photo   and
bremsstrahlung processes are those of Itoh et al.  (1996).  For plasma
processes   we   included  the   treatment   presented   in  Haft   et
al.  (1994).  Radiative  opacities  are  those  of  the  OPAL  project
(Iglesias  \&   Rogers  1996),   including  carbon  and   oxygen  rich
composition.   We  adopted  the  conductive opacities  of  Cassisi  et
al. (2007).  This prescription covers the whole  regime where electron
conduction  is relevant.   For the  high density  regime, we  used the
equation of state of Segretain  et al.  (1994), which accounts for all
the important contributions  for both the liquid and  solid phases ---
see Althaus et al. (2007a) and references therein. For the low-density
regime, we used  an updated version of the equation  of state of Magni
\& Mazzitelli (1979).  The release of latent heat upon crystallization
was   also   considered.   In   particular,   for  our   calculations,
crystallization sets in when the ion coupling constant reaches $\Gamma
=180$, where $\Gamma \equiv \langle Z^{5/3}\rangle e^2/a_e k_{\rm B}T$
and $a_e$ is the interelectronic distance. We considered a latent heat
release  of $k_{\rm  B}T$ per  ion, which  was spread  over  the range
$175<\Gamma<185$.   Convection was  treated  in the  formalism of  the
mixing length theory as given  by the ML2 parameterization (Tassoul et
al. 1990).

All our white  dwarf sequences were computed in  a consistent way with
the evolution of the chemical abundance distribution caused by element
diffusion along the whole cooling phase.  In particular, we considered
gravitational  settling and  chemical diffusion  of  $^4$He, $^{12}$C,
$^{13}$C, $^{14}$N  and $^{16}$O  --- see Althaus  et al.   (2003) for
details. In  our simulations, we  obtain high amounts of  $^{13}$C and
$^{14}$N at  the VLTP, in  agreement with observational  inferences in
the Sakurai object and, in the case of $^{14}$N, in some PG 1159 stars
(Werner \& Herwig 2006).  Our treatment of time-dependent diffusion is
based on the multicomponent gas treatment presented in Burgers (1969).
Diffusion velocities are evaluated at each evolutionary time step. For
the  white  dwarf  evolution,  is  worth mentioning  that  opacity  is
calculated taking into account the diffusion predictions for the heavy
element composition.

\subsection{Initial models}

The  initial  models  for  our  white dwarf  sequences  correspond  to
realistic  PG  1159  stellar  configurations  derived  from  the  full
evolutionary  calculations   of  their  progenitor   stars  for  solar
metallicity  (Miller Bertolami  \& Althaus  2006).  All  the sequences
were  computed  from  the   ZAMS  through  the  thermally-pulsing  and
mass-loss phases on the AGB  and finally to the born-again stage where
the remaining H is violently burnt.  After the born-again episode, the
H-deficient,   quiescent  He-burning   remnants  evolve   at  constant
luminosity  to the  domain of  PG1159  stars with  a surface  chemical
composition rich in He, carbon and oxygen (Miller Bertolami \& Althaus
2006).   This  new  generation  of  PG 1159  evolutionary  models  has
succeeded  in   explaining  both   the  spread  in   surface  chemical
composition observed in most PG 1159  stars and the location of the GW
Vir  instability  strip  in  the  $\log T_{\rm  eff}-  \log  g$  plane
(C\'orsico et al.   2006), as well as the  short born-again timescales
of V4334 Sgr (Miller Bertolami  et al.  2006).  To our knowledge, this
is the first  time that such a detailed  treatment of the evolutionary
history of progenitors stars  with different initial stellar masses is
taken  into  account  in  the  calculation of  cooling  sequences  for
H-deficient white dwarfs.

\begin{table*}
\centering
\caption{Main parameters associated to  the model with highest $T_{\rm
         eff}$ for  each sequence.  The tabulated abundances  (by mass
         fraction) correspond to the surface value.}
\begin{tabular}{ccccccc}
\hline     
\hline
$M/M_{\sun}$ & 
$\log T_{\rm eff}$ [K] & 
$\log(L/L_{\sun})$  & 
$\log g$ [cm~s$^{-2}$] & 
$X_{\rm He}$ & 
$X_{\rm C}$ & 
$X_{\rm O}$ \\ 
\hline
0.515 & 5.0634  &  2.6868  &  6.6713 & 0.7437 & 0.1637 & 0.0279  \\
0.530 & 5.1366  &  2.9762  &  6.6876 & 0.3301 & 0.3944 & 0.1716  \\
0.542 & 5.1650  &  3.0546  &  6.7412 & 0.2805 & 0.4064 & 0.2127  \\
0.565 & 5.2105  &  3.1565  &  6.8304 & 0.3632 & 0.2737 & 0.2173  \\
0.584 & 5.2398  &  3.2574  &  6.8615 & 0.3947 & 0.3060 & 0.1704  \\
0.609 & 5.2819  &  3.3533  &  6.9518 & 0.4990 & 0.3465 & 0.1033  \\
0.664 & 5.3578  &  3.3611  &  7.0813 & 0.4707 & 0.3260 & 0.1234  \\
0.741 & 5.4535  &  3.8069  &  7.2701 & 0.4795 & 0.3361 & 0.1390  \\
0.870 & 5.5829  &  4.0961  &  7.5679 & 0.5433 & 0.3012 & 0.0938  \\
\hline
\hline
\end{tabular}
\label{tablequi}
\end{table*}

Specifically,  we computed nine  H-deficient white  dwarf evolutionary
sequences. In Table  \ref{tableini} we list the stellar  masses of the
resulting  white  dwarfs,  together  with  the inital  masses  of  the
progenitor stars at the ZAMS.   Also listed in Table \ref{tableini} is
the  mass of  He  left by  previous  evolution.  Note  that for  these
sequences,  which cover  most of  the observed  stellar mass  range of
H-deficient  white dwarfs, the  expected range  of He  envelope masses
spans more than  one order of magnitude. Due to  He burning during the
PG  1159 stage,  the mass  of  He left  when the  remnant reaches  its
cooling track  is considerably lower than that  obtained shortly after
the born-again episode (0.0036 and $0.0075 \, M_{\sun}$, respectively,
for the $0.664 \, M_{\sun}$ sequence).  For the $1 \, M_{\sun}$ model,
two  different sequences  were computed  with two  different mass-loss
rates during  the AGB  evolution.  In this  way we obtain  a different
number of thermal pulses and, in the end, two different remnant masses
of 0.530 and $0.542\,  M_{\sun}$, respectively.  Finally, the sequence
with $0.515  \, M_{\sun}$ was  specifically computed to  reproduce the
observational features  of low-mass,  He-enriched PG 1159  stars, like
{\mbox{\object{MCT  0130$-$1937}}} and  {\mbox{\object{HS 1517+7403}}}
--- see Miller Bertolami \& Althaus (2006) and Althaus et al. (2007b).
This model was  extracted from the full evolution  of a $1\, M_{\sun}$
model star that experiences its  first thermal pulse as a late thermal
pulse (LTP) after leaving the AGB.  We adopt the mass-loss rates of of
Bl\"ocker  (1995) multiplied  by a  0.2 factor.  Mass loss  during the
departure  from the AGB  has been  arbitrarily set  to obtain  a final
helium shell  flash during  the early white  dwarf cooling  phase. Our
calculations cover the hot pre-white dwarf stages (PG 1159 regime) and
the advanced phases  of evolution down to an  effective temperature of
7000 K, therefore covering all the DB cooling sequence.

\begin{figure}
\begin{center}
\includegraphics[clip,width=0.9\columnwidth]{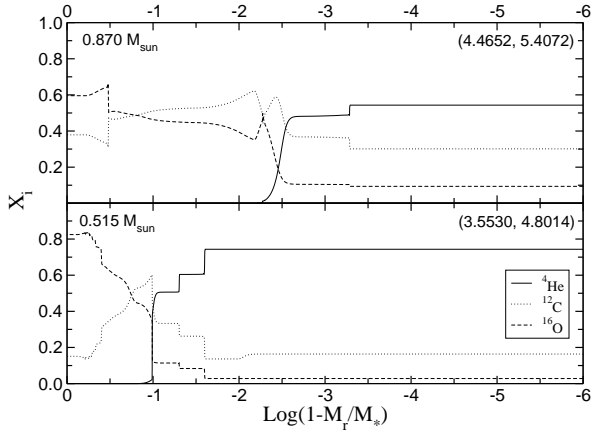}
\end{center}
\caption{Abundance  by  mass  of $^4$He,  $^{12}$C and  $^{16}$O  as a 
         function of the outer  mass fraction $\log(1-M_r/M_*)$ for PG
         1159  models corresponding  to  the sequences  of 0.870  (top
         panel)  and 0.515 $M_{\sun}$  (bottom panel).  Luminosity and
         effective  temperature ($\log(L/L_{\sun});\log  T_{\rm eff}$)
         are specified for each model.}
\label{input1}
\end{figure}

In Fig.  \ref{input1} we show  the mass abundances of $^4$He, $^{12}$C
and $^{16}$O throughout  the deep interior of two  PG 1159 star models
corresponding to the sequences with 0.515 and $0.870\, M_{\sun}$.  The
models shown  correspond to  evolutionary stages at  high luminosities
where the  chemical profile at the  bottom of the  He-rich envelope is
being modified  by residual He shell-burning.  Note  the dependence of
the composition profile on the stellar mass, which emphasizes the need
for a  detailed knowledge  of the progenitor  history for  a realistic
treatment of  white dwarf  evolution.  The chemical  stratification of
our  H-deficient pre-white  dwarf models  consists of  a carbon-oxygen
core, which  emerges from core He  burning in prior  stages.  The core
chemical  profile  is typical  of  situations  in  which extra  mixing
episodes beyond the fully convective  core during the core He burning,
like moderate overshooting and/or  semiconvection, are allowed --- see
Straniero et al. (2003) and  also Prada Moroni \& Straniero (2007) for
the consequences on white dwarf  evolution.  The core is surrounded by
a homogeneous  and extended  envelope rich in  He, carbon  and oxygen,
which  is the result  of prior  mixing and  burning events  during the
thermally pulsing AGB and born-again phases.  The surface abundance of
He of our pre-white dwarf models  is within 0.28 and 0.75 by mass (see
Table  \ref{tablequi}),  which  is  in  agreement with  the  range  of
observed He abundances  in most PG1159 stars (Werner  \& Herwig 2006).
These abundances are not only determined by the stellar mass, but also
by the  number of thermal  pulses during the  AGB phase, by  the inner
penetration of the convective envelope  and by the mass lost after the
complete  burning  of  protons  (Miller Bertolami  \&  Althaus  2006).
Finally, we have  removed any H from our PG  1159 models, despite that
some  H  is  expected  to  be  present in  PG  1159  stars.   This  is
particularly  true  if   some  of  these  stars  result   from  a  LTP
episode. Here,  the H content  with which the progenitor  star departs
from the  AGB for the first time  is not burnt during  the late helium
flash  but  instead diluted  to  deeper  layers.   In this  case,  the
presence of  H in PG 1159 stars  may be responsible for  the fact that
not   all   PG  1159   stars   evolve   to   DO  white   dwarfs,   see
Sect. \ref{conclusiones}.   By contrast, during the  VLTP, only traces
of  H  survive  the  violent   H  burning,  see  Miller  Bertolami  et
al. (2006). It  is expected that a large fraction  of this remaining H
is removed by mass loss during the following return to the giant state
and Sakurai state.


\section{Evolutionary results}\label{results}

\subsection{Sequences with thick envelopes}

\begin{figure}
\begin{center}
\includegraphics[clip,width=0.9\columnwidth]{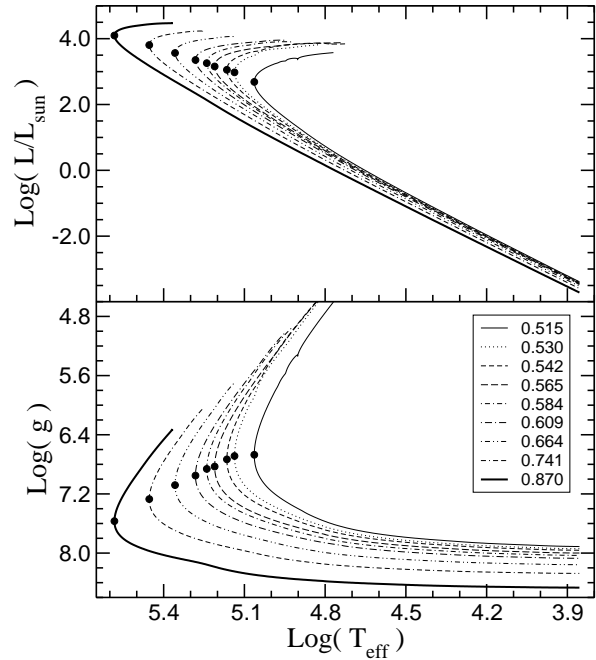}
\end{center}
\caption{Upper panel: Hertzsprung-Russell diagram for our evolutionary
         sequences.   Bottom  panel:  evolution  in the  $\log  T_{\rm
         eff}-\log g$  plane.  Filled dots on each  track indicate the
         instant  with  the highest  effective  temperature, which  we
         adopt  as our  time origin.  Mass values  are given  in solar
         units.}
\label{resultado1}
\end{figure}

In Fig.   \ref{resultado1}, we show  the evolution of  our H-deficient
sequences  both in the  Hertzsprung-Russell diagram  and in  the $\log
T_{\rm eff}--\log  g$ plane.   The evolutionary sequences  shown cover
from the  PG 1159 stage at  intermediate gravities to  the white dwarf
domain.  Thus, these sequences  constitute a homogeneous set of tracks
that    consistently   cover    both   evolutionary    stages.    Some
characteristics  of  them at  the  highest  effective temperature  are
listed in Table \ref{tablequi}.

\begin{figure}
\begin{center}
\includegraphics[clip,width=0.9\columnwidth]{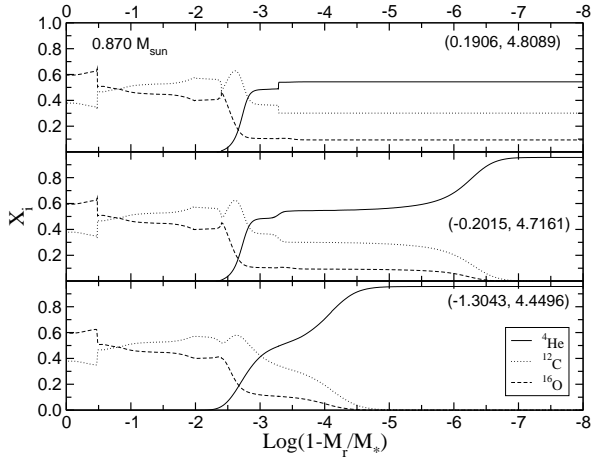}
\end{center}
\caption{Internal  abundance  distribution  of  $^4$He,  $^{12}$C  and
         $^{16}$O   as  a   function  of   the  outer   mass  fraction
         $\log(1-M_r/M_*)$  for our  $0.870 \,  M_{\sun}$  sequence at
         various selected  stages on its cooling  branch, as specified
         by   the   luminosity   and  effective   temperature   values
         ($\log(L/L_{\sun});\log T_{\rm eff}$) in each panel.}
\label{resultado-perfil-1}
\end{figure}

\begin{figure}
\begin{center}
\includegraphics[clip,width=0.9\columnwidth]{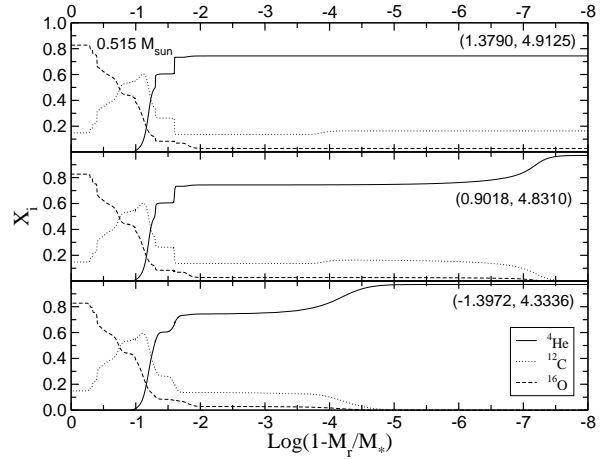}
\end{center}
\caption{Same as Fig. \ref{resultado-perfil-1}  but  for the $0.515 \, 
         M_{\sun}$ sequence.}
\label{resultado-perfil-2}
\end{figure}

During the PG 1159 and  hot white dwarf stages, $^{4}$He, $^{12}$C and
$^{16}$O  are by  far  the dominant  species  in the  envelope of  our
models.  Among the main  remaining constituents are $^{13}$C, $^{14}$N
and $^{22}$Ne.   In particular, as a  result of the  violent H burning
during the born-again episode,  $^{13}$C reaches abundances as high as
0.05  by mass  in the  envelope.   Once He  burning becomes  virtually
extinct  after  the  point   of  maximum  effective  temperature,  the
H-deficient  remnants  settle upon  their  cooling  track.  Here,  the
chemical  abundance  distribution will  be  strongly  modified by  the
various  diffusion processes  acting during  white dwarf  evolution. A
glimpse of the inner chemistry variations that occur during this stage
is      given     in      Figs.       \ref{resultado-perfil-1}     and
\ref{resultado-perfil-2}  for  the   0.870  and  $0.515  \,  M_{\sun}$
sequences,  respectively. The  upper panel  of each  figure  shows the
chemical stratification  at the start of the  cooling track.  Clearly,
gravitational settling causes  He to float to the  surface and heavier
elements to sink.   Thus, as the evolution proceeds,  the He abundance
in the  outer layers increases, thus  turning the PG 1159  star into a
white dwarf of  spectral type DO (Dehner \&  Kawaler 1995; Gautschy \&
Althaus 2002).  At  this point, we must mention  that our treatment of
diffusion cannot predict when a PG  1159 star is transformed into a DO
white dwarf. To  do this, a more elaborated  treatment of diffusion in
the surface layers than that  we consider here --- including mass loss
through winds, and radiative levitation --- should be used (Unglaub \&
Bues   2000).    Note   the   formation  of   a   diffusively-evolving
double-layered chemical  structure, that is,  a pure He mantle  and an
intermediate remnant shell rich in He, carbon and oxygen result of the
last  He thermal  pulse.  Because  of the  more  intense gravitational
field  and  the smaller  He  content  in  more massive  remnants,  the
double-layered  structure in  our $0.87  \, M_{\sun}$  sequence almost
disappears  before the  DBV instability  strip is  reached  at $T_{\rm
eff}\approx   30\,000$    K,   see   the   bottom    panel   of   Fig.
\ref{resultado-perfil-1}.  Indeed, the outer layer chemical profile of
the massive  remnant resembles that  of a single-layered  structure by
the time cooling has proceeded to this stage.  As shown by Fontaine \&
Brassard (2002) and also by  Althaus \& C\'orsico (2004), the presence
of a  double-layered structure bears consequences  for the pulsational
properties  of   variable  DB   white  dwarfs.   In   particular,  the
double-layered structure causes the period-spacing diagrams to exhibit
mode-trapping  substructure  (Althaus  \&  C\'orsico 2004).   Thus,  a
complete assessment  of the evolutionary history  of H-deficient white
dwarf progenitors  like that considered here constitutes  a key aspect
that has to be considered in asteroseismological studies of DBV stars.
In this  sense, we  conclude that on  the basis  of a full  account of
progenitor evolution,  a double-layered  structure is not  expected in
H-deficient  white  dwarfs with  masses  higher  than  about $0.87  \,
M_{\sun}$ by the time evolution has proceeded to the domain of the DBV
instability strip.

\begin{figure*}
\begin{center}
\includegraphics[clip,width=0.9\textwidth]{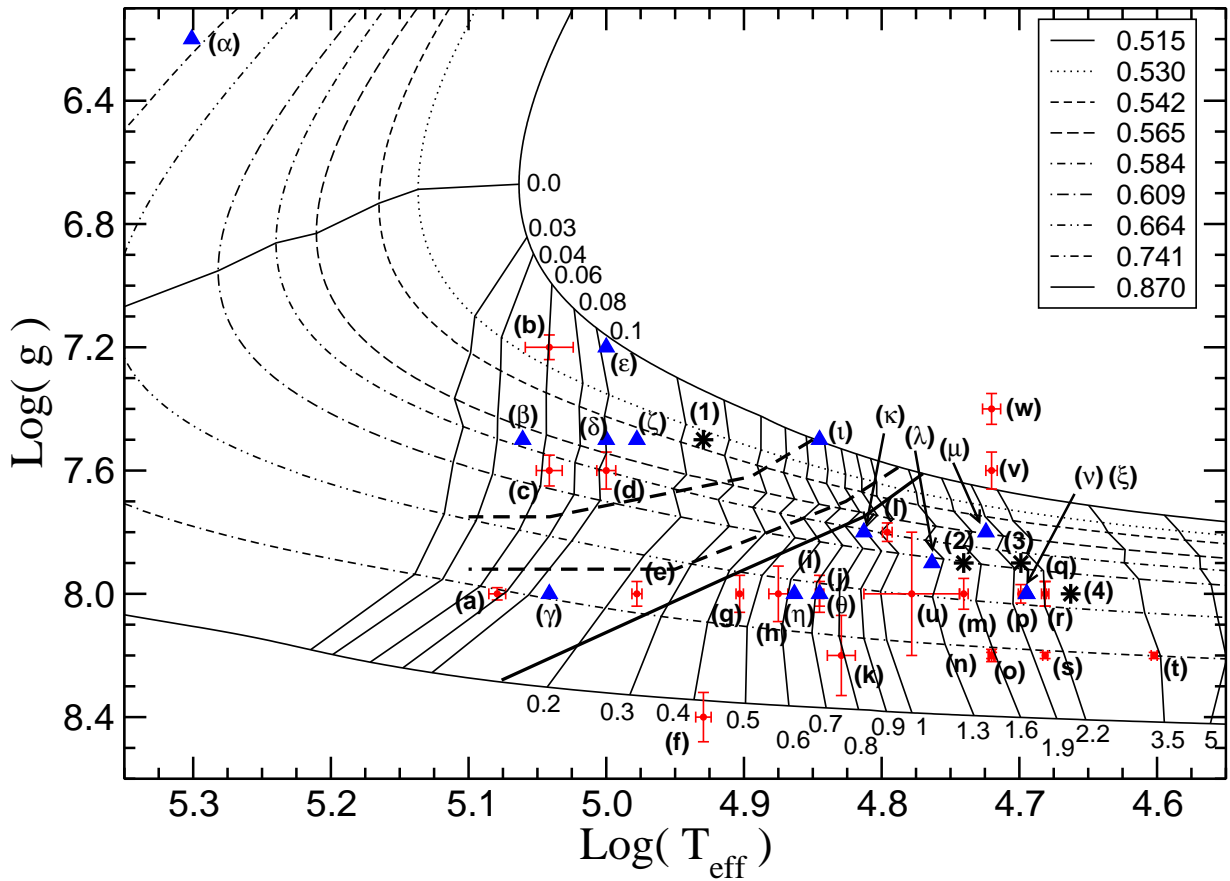}
\end{center}
\caption{Surface gravity $g$  (in cm/s$^2$) as a  function of  $T_{\rm  
         eff}$ for our H-deficient  sequences with thick He envelopes.
         From  bottom  to top,  curves  correspond  to sequences  with
         stellar  masses from  0.870 to  $0.515\, M_{\sun}$.   We also
         plot isochrones ranging from 0.03  to 5 Myr measured from the
         highest effective  temperature point.  The  observed DO white
         dwarfs analyzed  by H\"ugelmeyer  et al. (2006),  Dreizler \&
         Werner (1996), and Dreizler et al. (1997), as listed in Table
         \ref{tabledo}, are also  shown (filled circles and triangles,
         and  star  symbols,  respectively).   The  thick  solid  line
         corresponds to  the wind limit  for PG 1159 stars  taken from
         Unglaub \& Bues (2000), and the two thick dashed lines to the
         wind limits for  PG 1159 stars with H/He=0.1  and 0.01 (upper
         and lower line, respectively).}
\label{resultado2}
\end{figure*}

\begin{figure}
\begin{center}
\includegraphics[clip,width=0.9\columnwidth]{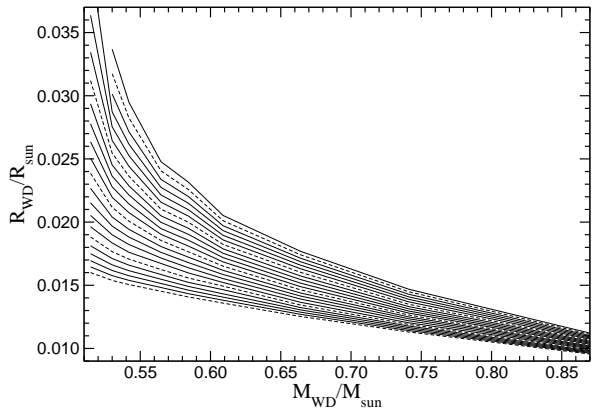}
\end{center}
\caption{Mass-radius  relations  for  our H-deficient  sequences  with
         thick  He  envelopes,   corresponding  to  several  effective
         temperatures. From bottom to  top we show relations for $40\,
         000$ K to $124\, 000$ K  in temperature steps of $4\, 000$ K.
         In  the interest  of clarity,  the mass-radius  relations for
         effective temperatures $40\, 000$ K, $60\, 000$ K, $80\, 000$
         K, $100\, 000$ K and  $120\, 000$ K are displayed with dashed
         lines.}
\label{resultado4}
\end{figure}

Our evolutionary sequences are  appropriate for mass determinations of
H-deficient  white dwarfs, particularly  at the  early and  hot stages
when  white  dwarf  evolution   is  still  sensitive  to  the  initial
conditions.  In  Fig.  \ref{resultado2} we  show the evolution  of our
sequences in  the $\log g-log  T_{\rm eff}$ diagram and  the effective
temperatures   and  surface   gravities  (in   cm/s$^2$)   derived  by
H\"ugelmeyer et al. (2006), Dreizler  \& Werner (1996) and Dreizler et
al. (1997) of  the observed DO white dwarfs. They  are listed in Table
\ref{tabledo}  together with  the stellar  masses determined  by these
authors using  the evolutionary tracks of Wood  (1995) for carbon-core
white  dwarfs and  a He  layer mass  of 10$^{-4}  M_*$ (column  4). In
passing, we  note that the stellar mass  of J204158.98+000325.4 listed
in H\"ugelmeyer  et al.   (2006) is not  correct, it should  read 0.49
instead  of  $0.6\,  M_{\sun}$.   We  have  corrected  this  in  Table
\ref{tabledo}.  From Fig.  \ref{resultado4} it is clear the importance
of  finite-temperature effects during  the early  evolutionary stages,
particularly in the  case of low-mass sequences.  We  expect then that
the  stellar mass  for  hot  DOs with  low  gravities determined  from
cooling tracks that neglect the progenitor evolution will be different
when our new evolutionary  tracks are employed.  Thus, we redetermined
the stellar mass of the  DO white dwarfs listed in Table \ref{tabledo}
using our new sequences.  The results are given in column 5.  For most
DOs,  mass determinations from  both sets  of tracks  agree reasonably
well,  but  in  the  case  of  the hot  and  lowest-gravity  DOs,  our
evolutionary tracks lead to an increase in the derived stellar mass of
up to 10\% when compared  with the determinations based on the cooling
tracks   of  Wood  (1995).    We  must   caution  however   that  some
simplifications done in the calculations of Wood (1995), in particular
regarding the  chemical profile  and He envelope  mass of  the models,
prevents us  from a consistent and meaningful  comparison between both
sets  of  tracks.   For  instance,  the decrease  in  surface  gravity
resulting from the  assumption of a pure carbon core  in the models of
Wood (1995) is  in part compensated for by the  lower He envelope mass
that characterize these models,  as compared to our carbon-oxygen core
white  dwarf  models  that   are  characterized  by  more  massive  He
envelopes.    Finally,    note   that   for    the   low-gravity   DO,
$J075606.36+421613.0$,  we   derive  a  stellar  mass   of  $0.467  \,
M_{\sun}$.   Although this  value  results from  extrapolation of  our
tracks,  it  is  interesting  to  note  that  the  resulting  mass  is
substantially   higher  than   the  $0.42   \,  M_{\sun}$   quoted  by
H\"ugelmeyer et  al.  (2006).  The  new stellar mass seems  to suggest
that  the progenitor  of this  star could  probably be  a  He-SdO star
formed in a delayed He core flash (Miller Bertolami et al. 2008).

\begin{table*}
\scriptsize
\caption{Stellar  parameters   of  the  DO  white   dwarf  samples  of
         H\"ugelmeyer et  al.  (2006), Dreizler \&  Werner (1996), and
         Dreizler  et al.   (1997) ---  roman and  greek  letters, and
         numbers, respectively.   Column 4  gives the stellar  mass in
         solar  units as derived  by these  authors using  the cooling
         tracks of  Wood (1995).  Column  5 and 6  list, respectively,
         the stellar mass in solar units derived from our evolutionary
         tracks with thick and thin He envelopes. The next two columns
         give  the white  dwarf ages.   The effective  temperature and
         gravity  of  KPD0005+5106  have  been taken  from  Werner  et
         al. (2008b).  For  this star, the lower limit  to log $g$ has
         been adopted. The  zero point for the ages  is at the highest
         effective temperature.}
\begin{center}
\begin{tabular}{lrccccccc}
\hline     
\hline
DO Name  & $T_{\rm eff}$~~~~  &  $\log g$             & $M_{\rm WD}$ &  $M_{\rm WD}$ [$M_{\sun}$] & $M_{\rm WD}$ [$M_{\sun}$] & Age [Myr] & Age [Myr] & Ref \\ 
            &     [kK]~~~~      &   [cm~s$^{-2}$]       & (Wood)      & (Thick)  & (Thin)   & (Thick) & (Thin)  & \\
\hline
J091433.61+581238.1   & 120.0$\pm$1.7 & 8.00$\pm$0.02 & 0.75 & 0.750$\pm0.010$           & 0.740$\pm0.010$            & 0.076$\pm0.006$           & 0.074$\pm0.006$ & (a)\\         
J204158.98+000325.4   & 110.0$\pm$4.5 & 7.20$\pm$0.04 & 0.49 & 0.525$_{-0.002}^{+0.003}$ & 0.510$_{-0.005}^{+0.007}$  & 0.060$_{-0.010}^{+0.016}$ & 0.030$_{-0.005}^{+0.007}$  & (b)\\
J214223.32+111740.4   & 110.0$\pm$2.4 & 7.60$\pm$0.05 & 0.60 & 0.595$_{-0.011}^{+0.012}$ & 0.580$_{-0.013}^{+0.015}$  & 0.064$_{-0.007}^{+0.008}$ & 0.054$\pm0.006$ & (c)\\
J154752.33+423210.9   & 100.0$\pm$1.6 & 7.60$\pm$0.06 & 0.59 & 0.584$_{-0.019}^{+0.014}$ & 0.569$_{-0.016}^ {+0.017}$ & 0.107$_{-0.009}^{+0.008}$ & 0.088$\pm0.008$ & (d)\\
J151026.48+610656.9   &  95.0$\pm$0.8 & 8.00$\pm$0.04 & 0.71 & 0.715$\pm0.015$           & 0.708$\pm0.016$            & 0.232$\pm0.008$           & 0.233$\pm0.008$ & (e)\\        
J084008.72+325114.6   &  85.0$\pm$1.1 & 8.40$\pm$0.08 & 0.90 & 0.900$_{-0.045}^{+0.040}$ & 0.890$\pm0.040$            & 0.420$\pm0.015$           & 0.460$\pm0.020$ & (f)\\
J025403.75+005854.5   &  80.0$\pm$0.5 & 8.00$\pm$0.06 & 0.68 & 0.695$\pm0.025$           & 0.688$_{-0.021}^{+0.027}$  & 0.450$\pm0.010$           & 0.455$\pm0.010$ & (g)\\
J155356.81+483228.6   &  75.0$\pm$1.2 & 8.00$\pm$0.09 & 0.68 & 0.687$_{-0.037}^{+0.043}$ & 0.680$_{-0.038}^{+0.043}$  & 0.565$_{-0.030}^{+0.034}$ & 0.575$\pm0.035$ & (h)\\
J084223.14+375900.2   &  70.0$\pm$0.3 & 8.00$\pm$0.06 & 0.68 & 0.680$_{-0.026}^{+0.027}$ & 0.673$_{-0.026}^{+0.029}$  & 0.715$\pm0.010$           & 0.725$\pm0.010$ & (i)\\
J140409.96+045739.9   &  70.0$\pm$0.5 & 8.00$\pm$0.04 & 0.68 & 0.680$_{-0.018}^{+0.017}$ & 0.673$_{-0.018}^{+0.019}$  & 0.715$\pm0.015$           & 0.725$\pm0.015$ & (j)\\
J113631.50+591229.8   &  67.5$\pm$1.6 & 8.20$\pm$0.13 & 0.78 & 0.775$_{-0.065}^{+0.070}$ & 0.770$_{-0.057}^{+0.070}$  & 0.790$\pm0.050$           & 0.820$\pm0.050$ & (k)\\
J131724.75+000237.4   &  62.5$\pm$0.5 & 7.80$\pm$0.03 & 0.58 & 0.582$_{-0.012}^{+0.013}$ & 0.576$_{-0.011}^{+0.012}$  & 1.040$\pm0.020$           & 0.950$\pm0.030$ & (l)\\
J034101.39+005353.0   &  55.0$\pm$0.4 & 8.00$\pm$0.05 & 0.65 & 0.656$\pm0.024$           & 0.650$\pm0.025$            & 1.500$\pm0.030$           & 1.500$\pm0.030$ & (m)\\        
J003343.06+142251.5   &  52.5$\pm$0.2 & 8.20$\pm$0.01 & 0.76 & 0.755$\pm0.005$           & 0.753$\pm0.005$            & 1.570$\pm0.010$           & 1.580$\pm0.010$ & (n)\\         
J115218.69$-$024915.9 &  52.5$\pm$0.3 & 8.20$\pm$0.02 & 0.76 & 0.755$\pm0.012$           & 0.753$\pm0.011$            & 1.570$\pm0.020$           & 1.580$\pm0.020$ & (o)\\         
J034227.62$-$072213.2 &  50.0$\pm$0.2 & 8.00$\pm$0.03 & 0.65 & 0.647$_{-0.014}^{+0.015}$ & 0.643$\pm0.015$            & 1.970$\pm0.020$           & 1.950$\pm0.020$ & (p)\\
J081115.09+270621.8   &  48.0$\pm$0.3 & 8.00$\pm$0.04 & 0.64 & 0.645$\pm0.020$           & 0.640$\pm0.020$            & 2.210$\pm0.040$           & 2.190$\pm0.040$ & (q)\\        
J113609.59+484318.9   &  48.0$\pm$0.1 & 8.00$\pm$0.04 & 0.64 & 0.645$\pm0.020$           & 0.640$\pm0.020$            & 2.210$\pm0.020$           & 2.190$\pm0.020$ & (r)\\        
J081546.08+244603.3   &  48.0$\pm$0.2 & 8.20$\pm$0.01 & 0.76 & 0.750$\pm0.007$           & 0.747$_{-0.005}^{+0.006}$  & 2.010$\pm0.030$           & 2.000$\pm0.020$  & (s) \\
J015629.58+131744.7   &  40.0$\pm$0.2 & 8.20$\pm$0.01 & 0.75 & 0.741$\pm0.007$           & 0.739$_{-0.007}^{+0.006}$  & 3.400$\pm0.050$           & 3.270$\pm0.050$  & (t)\\
J091621.83+052119.2   &  60.0$\pm$5.0 & 8.00$\pm$0.20 & 0.66 & 0.664$_{-0.087}^{+0.101}$ & 0.659$_{-0.087}^{+0.103}$  & 1.170$_{-0.260}^{+0.300}$ & 1.170$_{-0.250}^{+0.300}$ & (u) \\
J133633.22$-$013116.5 &  52.5$\pm$0.5 & 7.60$\pm$0.06 & 0.48 & 0.500$\pm0.015$           & 0.485$_{-0.015}^{+0.020}$  & 2.270$\pm0.070$           & 2.400$\pm0.100$  & (v)\\
J075606.36+421613.0   &  52.5$\pm$0.8 & 7.40$\pm$0.05 & 0.42 & 0.467$_{-0.003}^{+0.005}$ & 0.445$\pm0.005$            & 2.900$\pm0.150$           & 3.200$\pm0.150$  & (w)\\
 & & & & & &\\
KPD0005+5106          & 200.          & 6.2           & ---  & 0.770                     & 0.700                      & $-$0.00153                & $-$0.00038  & ($\alpha$) \\
PG0038+199            & 115.          & 7.5           & 0.59 & 0.575                     & 0.562                      & 0.050                     & 0.040 & ($\beta$) \\
PG0109+111            & 110.          & 8.0           & 0.74 & 0.735                     & 0.730                      & 0.120                     & 0.120 & ($\gamma$) \\
PG1034+001            & 100.          & 7.5           & 0.56 & 0.554                     & 0.545                      & 0.100                     & 0.080 & ($\delta$) \\
HS1830+7209           & 100.          & 7.2           & 0.47 & 0.517                     & 0.500                      & 0.107                     & 0.052 & ($\varepsilon$) \\
PG0108+101            &  95.          & 7.5           & 0.54 & 0.548                     & 0.540                      & 0.133                     & 0.105 & ($\zeta$) \\
PG0046+078            &  73.          & 8.0           & 0.68 & 0.680                     & 0.675                      & 0.620                     & 0.630 & ($\eta$) \\
PG0237+116            &  70.          & 8.0           & 0.68 & 0.680                     & 0.675                      & 0.720                     & 0.730 & ($\theta$) \\
RE0503-289            &  70.          & 7.5           & 0.49 & 0.515                     & 0.505                      & 0.670                     & 0.700 & ($\iota$) \\ 
HS0111+0012           &  65.          & 7.8           & 0.58 & 0.587                     & 0.580                      & 0.880                     & 0.820 & ($\kappa$) \\ 
Lanning 14            &  58.          & 7.9           & 0.61 & 0.615                     & 0.610                      & 1.220                     & 1.225 & ($\lambda$) \\ 
HZ21                  &  53.          & 7.8           & 0.56 & 0.562                     & 0.555                      & 1.750                     & 1.750 & ($\mu$) \\ 
HS0103+2947           &  49.5         & 8.0           & 0.65 & 0.645                     & 0.640                      & 2.025                     & 2.010 & ($\nu$) \\ 
HD149499B             &  49.5         & 8.0           & 0.65 & 0.645                     & 0.640                      & 2.025                     & 2.010 & ($\nu$) \\ 
& & & & & &\\                                                                                                                                   
MCT 2148-294          &  85.          & 7.5           & 0.53 & 0.536                     & 0.529                      & 0.240                     & 0.210 & (1)  \\
PG 0929+270           &  55.          & 7.9           & 0.61 & 0.608                     & 0.603                      & 1.450                     & 1.460 & (2)  \\
PG 1057-059           &  50.          & 7.9           & 0.60 & 0.601                     & 0.595                      & 2.025                     & 1.980 & (3)  \\ 
PG 1133+489           &  46.          & 8.0           & 0.64 & 0.640                     & 0.635                      & 2.500                     & 2.470 & (4)  \\

\hline
\hline
\end{tabular}
\end{center}
\label{tabledo}
\end{table*}

Our sequences  provide also realistic  evolutionary ages, particularly
for  the hot  stages of  white  dwarf evolution,  for which  employing
artificial procedures to  generate starting white dwarf configurations
may  lead to  evolutionary sequences  with  erroneous thermomechanical
structures and, thus, to wrong ages.  We employed our tracks to derive
evolutionary   ages   for   the   white   dwarfs   plotted   in   Fig.
\ref{resultado2}. They are listed  in column 7 of Table \ref{tabledo}.
The zero-point  age corresponds  to the highest  effective temperature
for each track, see Fig. \ref{resultado1}.  For the stellar masses not
covered by our tracks, ages  were obtained from extrapolation, so they
should be  taken with some caution.   We also plot  isochrones in Fig.
\ref{resultado2}  for better  visualization.   Note that  the bulk  of
observed DO white dwarfs have ages between 0.6 and 2.5 Myr, though the
hottest members  of the sample are  indeed very young,  with ages less
than 0.10 Myr.  Note also that the mass distribution of young DO white
dwarfs differs considerably from that  of the older ones. In fact, for
ages less than about $\sim 0.4$ Myr, DO white dwarfs less massive than
$\approx0.6\, M_{\sun}$ are  75\% of the total number  of DOs at those
ages,  whilst for  longer ages,  this contribution  declines  to 40\%.
Finally, the population of high-gravity DOs increases considerably for
ages  greater than  0.4 Myr,  below $T_{\rm  eff} \approx  85\,000$ K.
These  features  may   help  to  shed  light  on   the  formation  and
evolutionary status of DOs.

\begin{figure}
\begin{center}
\includegraphics[clip,width=0.9\columnwidth]{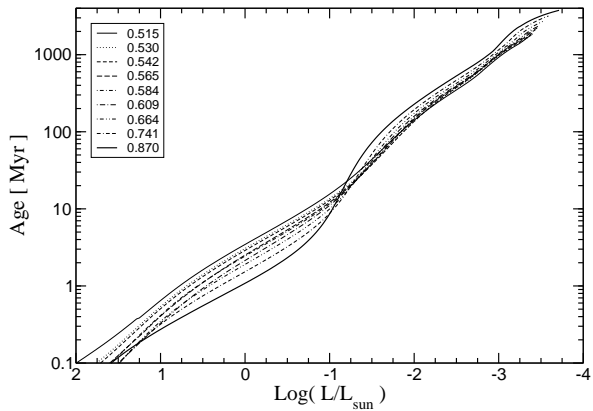}
\end{center}
\caption{Age  (in Myr)  versus  luminosity  (in  solar  units) for our
         H-deficient white dwarf sequences.}
\label{resultado3}
\end{figure}

In Fig.  \ref{resultado3}  we show the age (Myr) as  a function of the
luminosity  (in solar units)  for all  the H-deficient  sequences.  At
very high luminosity stages,  residual He shell-burning constitutes an
appreciable fraction of the surface luminosity. For instance, at $\log
T_{\rm eff}=  5$, the He  shell burning luminosity represents  25\% of
the surface luminosity in our lowest-mass sequence.  This contribution
declines steeply to 10\% at $\log T_{\rm eff}= 4.95$. Except for these
very  high luminosity  stages, evolution  during the  early  stages is
driven by  neutrino emission and  the release of  gravothermal energy.
In particular, neutrino losses exceed photon luminosity during most of
the hot white dwarf stages. Indeed, except for the hot and low-gravity
DO  sample  shown  in   Fig.   \ref{resultado2},  it  is  during  this
``neutrino epoch'' that most DO  white dwarfs are observed.  Note from
Fig.  \ref{resultado3}  that the imprints of neutrino  emission on the
cooling curve are more noticeable  as the mass increases. Less massive
sequences are older than  more massive sequences for $\log(L/L_{\sun})
\gtrsim -1$.   At this  stage, neutrino emission  arrives at  its end,
thus  causing  a change  in  the slope  in  the  cooling curve  around
$\log(L/L_{\sun})\sim -1$,  and later for the  less massive sequences.
At  lower luminosities,  Coulomb interactions  become  more important,
increasing the specific heat, until crystallization and the associated
release of latent  heat start at the center (Van  Horn 1968; Winget et
al. 2009).  At this point, cooling can be well understood on the basis
of the  simple cooling  theory of Mestel  (1952), which  predicts less
massive white dwarfs to cool faster.

\begin{figure}
\begin{center}
\includegraphics[clip,width=0.9\columnwidth]{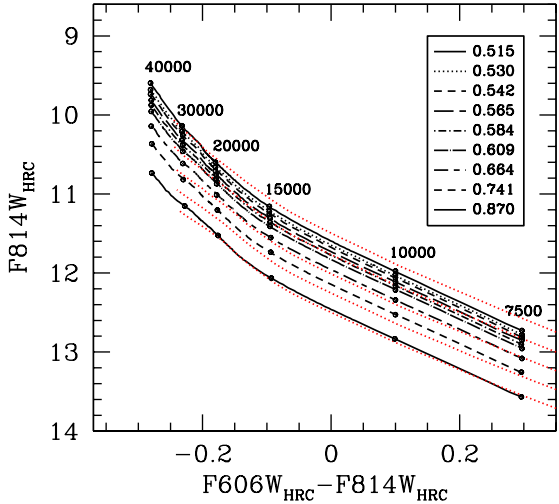}
\end{center}
\caption{F814W versus F606W--F814W  color-magnitude diagram for our DB
         white dwarf  sequences with  pure He envelopes.  Dotted lines
         correspond   to  Bergeron   et  al.   (1995)   sequences  for
         He-atmospheres white  dwarfs and stellar masses  of, from top
         to bottom, 0.5, 0.6, 0.7, 0.8, and $0.9\, M_{\sun}$. }
\label{cmd}
\end{figure}

For  all  our  cooling  sequences,  we computed  accurate  colors  and
magnitudes  based  on  non-gray  LTE model  atmospheres  (Rohrmann  et
al. 2002).  The  calculations were done for a  pure helium composition
and for the HST ACS  filters (Vega-mag system) and $UBVRI$ photometry.
Briefly,   each  model   atmosphere  is   approximated,   locally,  by
plane-parallel   layers   in   hydrostatic   equilibrium   and   local
thermodynamic  equilibrium.   Convective  energy transport  in  cooler
atmospheres   has    been   calculated   with    the   so-called   ML2
parameterization  of  the  mixing-length  theory.  The  correct  total
energy flux transported by radiation and convection is obtained from a
Rybicki  scheme, see  details in  Rohrmann et  al.   (2002).  Chemical
populations are computed within the occupation probability formalism
of Hummer  \& Mihalas (1988).   For helium pure  models, the  opacity sources
include  bound-free   (He,  He$^+$,  He$^{2+}$)   and  free-free  (He,
He$^+$,He$^{2+}$,  He$^-$) transitions,  electronic and  Rayleigh (He)
scattering, and  the most  significant line series  of He  and He$^+$.
The  level occupation  probabilities  are explicitly  included in  the
evaluation of the line and  continuum opacities.  As an example of the
present atmosphere calculations, we show in Fig.  \ref{cmd} the F814W,
F606W--F814W  color-magnitude diagram for  all our  sequences.  Colors
have  been computed  for effective  temperatures lower  than 40\,000K,
because  NLTE effects  become important  above this  temperature.  For
higher  effective  temperatures, most  of  radiation  flux emerges  at
shorter wavelengths.  We then expect colors to present a markedly less
dependence with  temperature.  Comparison between  our color sequences
and those due  to Bergeron (see Fig.  \ref{cmd})  reveals a rather good
agreement, with some small discrepancies presumably due to differences
of the constitutive physics used in the atmosphere codes.

\subsection{Thin envelope sequences}

In  section 3.1  we studied  sequences characterized  by  the thickest
He-rich envelope that can be left after a born-again episode. However,
there is theoretical and observational evidence that He-rich envelopes
could be substantially thinner.  For instance, using asteroseismology,
Kawaler \& Bradley (1994) derived  a He-rich envelope of $\sim 10^{-3}
M_{\sun}$ in  at least one  of the PG  1159 stars with masses  $0.6 \,
M_{\sun}$.  More recently, Althaus et  al.  (2008) showed that PG 1159
star models with  thin He-rich envelopes appear to  be needed to solve
the discrepancy  between the observed  (Costa \& Kepler 2008)  and the
theoretical rates of period change in the pulsating star PG1159$-$035.
A more direct evidence for the existence of PG 1159 stars with thin He
envelopes is provided  by the strong mass-loss events  reported in the
post-born again  star V4334 Sgr,  also known as Sakurai's  object (van
Hoof et  al.  2007).   All these facts  point at the  possibility that
some PG 1159 stars could be characterized by He-rich envelopes thinner
than expected  from the standard theory of  stellar evolution (Althaus
et al.  2009a).  Last but not  least, the mere existence of the hot DQ
class of white  dwarfs (Dufour et al. 2007,  2008) could be reflecting
the existence of He-rich white dwarfs with extremely thin He envelopes
(Dufour  et al.   2008;  Althaus et  al.   2009b).  In  view of  these
considerations, we examined the  evolution of H-deficient white dwarfs
for  the extreme  situation in  which a  very thin  He  envelope could
result from prior evolution.  To  this end, we computed five sequences
with  stellar  masses of  0.53,  0.565,  0.609,  0.664 and  $0.741  \,
M_{\sun}$ characterized by He-rich envelopes with masses $\sim 10^{-8}
M_{*}$,   a  value   which   is   of  the   order   required  by   the
diffusive/convective mixing  scenario to explain the origin  of hot DQ
white  dwarfs (Althaus et  al.  2009b).   To construct  initial models
with  such thin  He-rich  envelopes, we  artificially  reduced the  He
content of our  models to $\sim 10^{-8} M_{*}$  well before they reach
the PG 1159 domain.  These sequences were evolved down to $T_{\rm eff}
\approx$ 20\,000 K, where the He-rich envelope is diluted by the outer
convection zone.

\begin{figure}
\begin{center}
\includegraphics[clip,width=0.9\columnwidth]{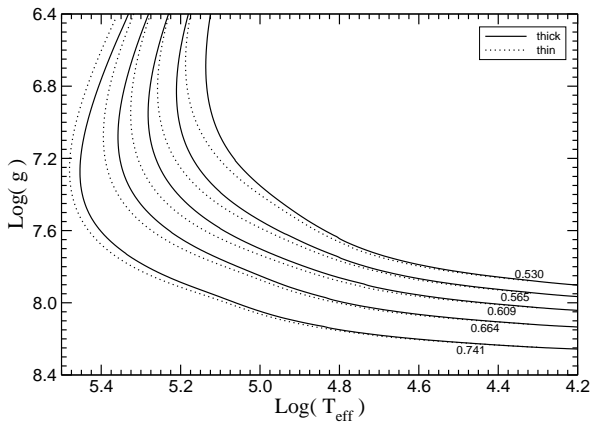}
\end{center}
\caption{Surface  gravity  as  a  function  of  $T_{\rm eff}$  for the 
         sequences with thin and thick envelopes and stellar masses of
         0.530, 0.565, 0.609, 0.664 and $0.741 \, M_{\sun}$.}
\label{resultado6}
\end{figure}

\begin{figure}
\begin{center}
\includegraphics[clip,width=0.9\columnwidth]{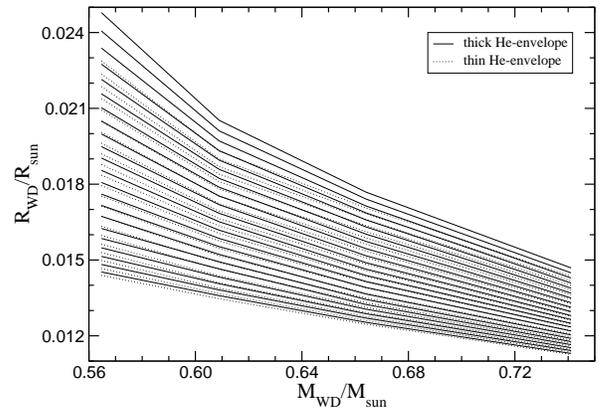}
\end{center}
\caption{Mass-radius relations for  H-deficient  sequences with  thick 
         (solid  lines) and  thin (dotted  line) He  envelopes. Curves
         correspond to effective temperatures of, from  bottom to top, 
         $40\,000$ K to $124\,000$ K in  temperature steps of $4\,000$
         K.}
\label{resultado7}
\end{figure}

\begin{figure*}
\begin{center}
\includegraphics[clip,width=0.9\textwidth]{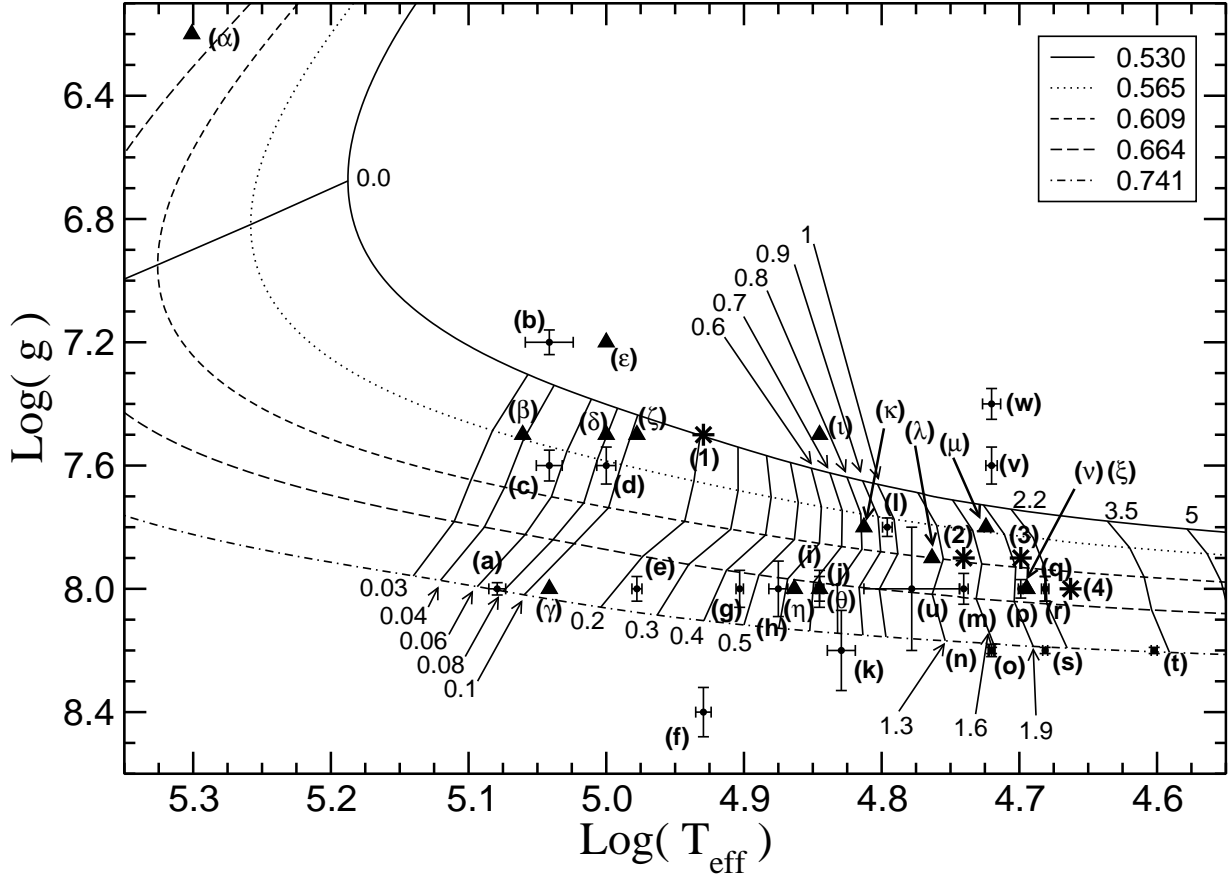}
\end{center}
\caption{Same as  Fig. \ref{resultado2} for our H-deficient  sequences 
         with thin He envelopes. We also plot isochrones (Myr) ranging
         from 0.03  to 5  Myr. The location  of the observed  DO white
         dwarfs listed in Table \ref{tabledo} is also shown.}
\label{resultado5}
\end{figure*}

In Fig.  \ref{resultado6}  the resulting tracks in the  $\log g - \log
T_{\rm eff}$  plane are compared  with the sequences  corresponding to
thick He  envelope models analyzed  in section 3.1. When  thin He-rich
envelopes are  considered, the evolutionary  tracks in this  plane are
strongly modified for the  PG 1159 regime, with important consequences
for PG  1159 spectroscopic mass  determinations (Althaus et  al 2008).
For the  white dwarf  regime, models with  thin He envelopes  are more
compact   ---  see   Fig.    \ref{resultado7}  ---   and,  thus,   are
characterized  by higher gravities.   As a  result, lower  white dwarf
masses would be inferred when the sequences with thin He envelopes are
used, as  it is  clear by examining  column 6 in  Table \ref{tabledo}.
From Fig.   \ref{resultado6} we expect  this to be relevant  only for
the  hot white dwarf  stages.  Indeed,  note from  Table \ref{tabledo}
that only for hot DO white  dwarfs we find a noticeable, albeit small,
change in the stellar mass when the models with thin He-rich envelopes
are used. In Fig.  \ref{resultado5} we show the white dwarf tracks for
thin He-rich envelopes  together with the observed DO  white dwarfs as
listed in Table \ref{tabledo}.   Note that the low-mass sequences with
thin He  envelopes evolve faster  during the hot stages  of evolution.
In  fact, for  these DO  white dwarfs  significantly smaller  ages are
inferred when the tracks with thin He envelopes are used, see column 8
in Table \ref{tabledo}.  This is because the He shell burning does not
represent an  energy source in  the sequences with thin  He envelopes.
Finally, the  formation of a double-layered structure  in pulsating DB
stars  is not  expected if  H-deficient white  dwarfs are  formed with
sufficiently thin He-rich envelopes.


\section{Discussion and conclusions}
\label{conclusiones}

Motivated by  the increasing number of detected  hot H-deficient white
dwarfs (of the  DO type) as well  as by the lack of  realistic sets of
evolutionary tracks appropriate  for their interpretation, we computed
new  evolutionary sequences and  colors for  non-DA white  dwarfs with
stellar  masses of 0.515,  0.530, 0.542,  0.565, 0.584,  0.609, 0.664,
0.741 and $0.870 \, M_{\sun}$.  In contrast to previous studies, these
sequences  were  obtained  considering   the  full  evolution  of  the
progenitor  stars, from  the ZAMS  through the  thermally  pulsing AGB
phase and born-again  episode, to the domain of the  PG 1159 stars ---
see Miller  Bertolami \& Althaus (2006) for  details.  Our H-deficient
white   dwarf   evolutionary  tracks   are   thus   not  affected   by
inconsistencies  arising from  artificial procedures  to  generate the
starting white dwarf configurations,  and so, they are appropriate for
realistic  mass and  age determinations  of H-deficient  white dwarfs,
particularly for  objects at the  early and hot  stages.  Calculations
were followed  to advanced  phases of evolution  down to  an effective
temperature of 7\,000 K.

We applied this new set of tracks to redetermine the stellar masses of
all known DO  white dwarfs with spectroscopically-determined effective
temperatures  and   gravities.   The  results  are   listed  in  Table
\ref{tabledo}.   The  stellar  masses  for  hot DOs  with  the  lowest
gravities  inferred  from  our  tracks  differ  appreciably  from  the
determinations based  on the evolutionary tracks of  Wood (1995). This
is  because  these  tracks  do  not consider  the  full  evolution  of
progenitor stars.   This is  also true for  the age  determinations of
these  stars.   In  particular,  our  evolutionary  sequences  provide
realistic  ages at  very  early  stages of  evolution.   For thick  He
layers, we find that He shell-burning supplies an appreciable fraction
of the surface luminosity,  thus changing appreciably the evolutionary
time  scales, particularly  of the  hottest DO  white dwarfs  with low
gravities. The  amount of He left  in the remnant  white dwarf affects
both the mass and the age determinations of hot DOs.  Models with thin
He-rich envelopes  are characterized  by higher gravities  than models
with thick He-rich envelopes, thus predicting lower white dwarf stellar
masses for the observed DOs.

\begin{figure}
\begin{center}
\includegraphics[clip,width=0.9\columnwidth]{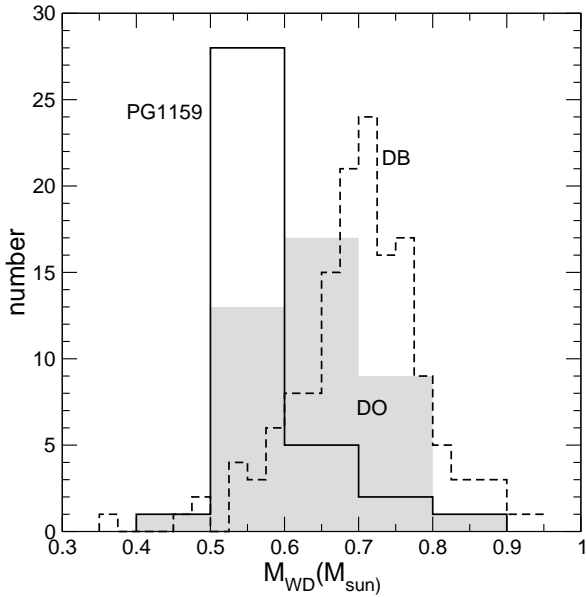}
\end{center}
\caption{The shaded  histogram shows the mass distribution  for the DO
         white dwarfs listed in  Table 3. The mass determinations have
         been  done using  the evolutionary  tracks described  in this
         work.   The  solid  line  corresponds  to the  PG  1159  mass
         distribution from Miller Bertolami \& Althaus (2006), and the
         dashed line to  that of DB white dwarfs  taken from Kepler et
         al. (2007).}
\label{histo}
\end{figure}

The evolutionary tracks  presented here cover all the  stages from the
domain of luminous PG 1159 star  region to the hot white dwarf domain.
Thus,  they constitute an  homogeneous set  of evolutionary  tracks to
study both  type of stars.  This  is of major interest  because we can
make consistent mass determinations for both the DOs and PG 1159 stars
for  the  first  time.   In  Fig.  \ref{histo}  we  compare  the  mass
distribution of the DO white dwarfs listed in Table \ref{tabledo} with
the mass  distribution of  PG 1159 stars  constructed on the  basis of
mass determinations made by Miller Bertolami \& Althaus (2006), shaded
region  and solid line,  respectively. The  size of  the bins  in both
distributions  is $0.1\,  M_{\sun}$. Note  that massive  DOs outnumber
massive  PG  1159 stars.   We  obtain  a mean  DO  mass  of $0.644  \,
M_{\sun}$, considerably higher (by  $0.071 \, M_{\sun}$) than the mean
mass  of PG 1159  stars, $0.573\,  M_{\sun}$.  One  may wonder  if the
reason for  this difference  is simply that  evolution through  the PG
1159  stage proceeds  considerably  faster for  massive stars  (Miller
Bertolami  \&  Althaus  2006),  thus  decreasing  the  possibility  of
detection.  To elaborate on this  point, we have proceeded as follows.
In order  to avoid the possibility  that the massive  PG1159 stars are
under-represented because they do not  spend long times in the PG 1159
stage, we have  tested the hypothesis that PG 1159  stars and DO white
dwarfs  form an  isolated evolutionary  sequence by  analyzing  the PG
1159+DO group as  a whole.  Taking advantage of  the consistent set of
tracks for both groups, we have divided the sample (78 stars) into two
groups  based on their  age. The  young group,  which consists  of all
stars with ages below 0.3 Myr (this includes all 37 PG 1159 stars plus
12 young DO white dwarfs) and the old group consisting of all DO white
dwarfs with  ages greater than  0.4 Myr (29  stars). It must  be noted
that there are no PG 1159 stars in the old group.

The  choice of the  0.3--0.4 Myr  isochrone to  divide both  groups is
admittedly arbitrary  but suggested by  the location of  the different
wind limits studied by Unglaub  \& Bues (2000).  Dividing the whole PG
1159+DO group  by an isochrone allows  to test the  hypothesis that PG
1159 and DO stars form an isolated evolutionary sequence.  This can be
done comparing their mass  distributions (as relative numbers within a
given  group  will not  be  affected  by  differences in  evolutionary
speeds).  The  two groups show  significant differences in  their mass
distributions.  While the  young  group  has a  mean  mass of  $M_{\rm
Y}=0.584  \, M_{\sun}$ (with  a variance  of $\sigma_{\rm  Y}=0.080 \,
M_{\sun}$)  the old  group  has a  mean  mass of  $M_{\rm O}=0.655  \,
M_{\sun}$  (with a  variance of  $\sigma_{\rm O}=0.087  \, M_{\sun}$).
Thus,  the  difference in  the  mean  masses  is comparable  with  the
variance within  each sample, suggesting that  both mass distributions
are    significantly   different.     This   is    confirmed    by   a
Kolmogorov-Smirnov  test.   The  result  of  this  test  is  that  the
probability  that   both  samples  are   taken  from  the   same  mass
distribution is rather small ($\sim 10^{-5}$).

These results clearly  suggest that there may be  some other important
evolutionary  channels operating  within PG  1159 and  DO  stars.  For
instance, it is possible that not all PG 1159 stars evolve to DOs.  In
particular,  those PG  1159 stars  resulting from  a LTP  --- low-mass
remnants are more  prone to experience a LTP  episode --- are expected
to evolve  into DA white dwarfs  (near the winds limits  shown in Fig.
\ref{resultado2} with solid dashed lines), thus avoiding the DO stage,
as shown by  Unglaub \& Bues (2000).  This is  so because, in contrast
to a VLTP event, in a LTP, H is not burned, but instead diluted to low
surface abundances (Miller  Bertolami \& Althaus 2006). Alternatively,
some DO white dwarfs could  result from evolutionary channels that not
involve only the PG 1159 stars. For instance, they could be the result
of post-merger evolution involving  the giant, H-deficient RCrB stars,
via the evolutionary link  RCrB\ $\rightarrow$ EHe (extreme He stars)\
$\rightarrow$ He-SdO$^+$  $\rightarrow$O(He) $\rightarrow$ DO  --- see
Rauch  et al. (2006)  for a  connection between  O(He) stars  and RCrB
stars.   This is  reinforced  by the  recent  study by  Werner et  al.
(2008a,b) of the star KPD0005+5106,  the hottest known DO white dwarf.
These authors  present evidence that KPD0005+5106 is  not a descendant
of PG  1159 stars, but  more probably related  to the O(He)  stars and
RCrB stars.

The  group  of   young  and  hot  DO  white   dwarfs  is  particularly
relevant. Note that they are located considerably above the wind limit
for PG 1159  stars (see Fig.  \ref{resultado2}).  In  fact, above this
line,   mass-loss   is   large   enough   to   prevent   gravitational
settling. Hence, the transformation of a  PG 1159 star into a DO white
dwarf should be expected  approximately below this line.  Although the
estimated mass-loss rates could be overestimated by more than a factor
of ten ---  in which case the  transition from PG 1159 to  DO would be
near a  line with $\log  g\approx 7.5$ (Unglaub  \& Bues 2000)  --- it
could be  possible that  the less  massive and young  DOs are  not the
descendants of PG 1159 stars.  This is also suggested by the fact that
for ages less  than about 0.3--0.4 Myr, the  DO white dwarf population
is markedly less  massive than at greater ages.  Indeed, the number of 
detected old, less massive DOs is
smaller  than the  number  of  young, less  massive  DOs, despite  the
evolutionary timescales being  a factor about 5 for  the former.  This
could be  indicative that  some of the  less massive DOs  experience a
tranformation in  their spectral type  after $\approx 0.3-0.4$  Myr of
evolution.   In particular,  traces of  H could  turn the  white dwarf
spectral type from DO to DA as a result of gravitational settling.  An
approximate  location  where   this  transformation  should  occur  is
provided by  the thick dashed  lines shown in  Fig.  \ref{resultado2},
which  provides the  wind  limits  for PG  1159  stars with  different
initial H abundances  (Unglaub \& Bues 2000).  It  is worth mentioning
that because  of the much  smaller carbon content  and correspondingly
less intense winds,  such limits would be expected  at lower gravities
in a DO  white dwarf with similar H content.   According to Unglaub \&
Bues (1998), for a DO white dwarf  with $\log g= 7$ and H number ratio
of 0.01, the transformation into a  DAO in presence of a moderate mass
loss would take about 0.25 Myr,  not very different from the age range
where the  DO mass distribution seems  to change. At this  point it is
worthwhile to comment on the  possibility that the origin of low-mass,
hot  DOs can be  connected with  the merger  of two  low-mass, He-core
white dwarfs  (Guerrero et al.  2004; Lor\'en--Aguilar  et al.  2009),
which will evolve  to become a low-gravity extreme He  star and then a
hot subdwarf  (Saio \&  Jeffery 2000). The  presence of H  expected in
this case, could lead to a  transformation of the DO white dwarfs into
DA ones.

In Fig.  \ref{histo}  we have also included with  dashed line the mass
distribution of the DB white dwarfs in SDSS hotter than $T_{\rm eff} =
16\,000$ K  taken from  Kepler et al.  (2007). The DB  distribution is
more  weigthed to  massive stars.  In  fact, the  DB mean  mass is  of
$0.711\,  M_{\sun}$, considerably  higher  than the  mean  DO mass  of
$0.644 \, M_{\sun}$.  This  marked difference in the mass distribution
between both populations is certainly  quite unexpected in view of our
understanding that DO stars must  evolve into DB stars. However, it is
possible that  the DB stellar  masses considered in  Fig.  \ref{histo}
could be overestimated.   This hypothesis was raised by  Kepler et al.
(2007), who discuss  that an increase in the  measured surface gravity
of DBs below $T_{\rm eff} \approx 18\,000$ K could result from missing
or incorrect  physics in model  atmospheres at such  temperatures.  It
could be also possible that the  increase in the gravity of the DBs by
$T_{\rm eff}\approx  18\,000$ K could  be reflecting the  existence of
DBs with traces of carbon in  their atmospheres. As shown by Dufour et
al.  (2005) in the context  of cool DQs, the surface gravities derived
from model  atmospheres including carbon are  significantly lower than
the gravities obtained from pure He models.  The idea that some carbon
could be  present in  the atmosphere of  some hot DBs  is particularly
attractive if a  large fraction of DB white  dwarfs evolve through the
hot DQ  stage. Indeed, as  discussed in Althaus  et al (2009b)  in the
diffusive/convective mixing picture for  the hot DQ formation, shortly
after the formation of the carbon-rich convective envelope, by $T_{\rm
eff} \approx 20\,000$ K, a substantial depletion of carbon is expected
in the entire convective envelope, because of the very short diffusion
timescale at the  base of the convection zone.   Hence, it is expected
that the hot  DQ stage, during which the  white dwarf is characterized
by  high amounts of  carbon in  its surface,  is indeed  a short-lived
phase.  However, the change  in composition in the convective envelope
will lead to  changes in the depth of the  convection zone, which will
affect, in  turn, the timescales  of diffusion.  The lack  of detailed
computations of this feedback between convection and diffusion shortly
after  the  formation  of a  DQ  does  not  allow to  make  reasonable
predictions,  but  it  is  not  discarded that  the  star  experiences
additional  mixing  episodes,   eventually  reaching  some  stationary
situation in  which the star resembles  a He-rich DB  white dwarf with
some carbon in its atmosphere.   This being the true course of events,
it would be worthwhile to re-determine the mass distribution of DBs in
the frame  of model  atmospheres that incorporate  the effects  of the
presence of  carbon.  Finally,  in view of  the material  discussed in
this  section,  it  is  feasible  that  a  fraction  of  low-mass  DOs
experiences a change in their spectral type at young ages.

In closing, because our DO tracks consider the evolutionary history of
progenitor, we  can make sound predictions about  the chemical profile
expected  in the  envelope of  H-deficient white  dwarfs.  This  is of
relevance for asteroseismological  inferences about pulsating DB white
dwarfs.  In  particular, we find  that depending on the  stellar mass,
diffusion processes  lead to a variety of  envelope chemical profiles,
from  single-layered   to  double-layered  structures,   by  the  time
evolution has  proceeded to the  domain of the variable  DBs. Detailed
tabulations  of our evolutionary  sequences are  available at  our web
site {\tt http://www.fcaglp.unlp.edu.ar/evolgroup}.


\acknowledgements This research was  supported by AGENCIA: Programa de
Modernizaci\'on Tecnol\'ogica  BID 1728/OC-AR, by the  AGAUR, by MCINN
grant AYA2008--04211--C02--01,  by the European Union  FEDER funds and
by PIP 6521 grant from CONICET.   LGA also acknowledges a PIV grant of
the AGAUR  of the  Generalitat de Catalunya.  We also  acknowledge the
comments  and suggestions of  our referee,  K. Werner,  which strongly
improved the original version of this work. This research has made use
of NASA's Astrophysics Data System.  Finally, we thank H.  Viturro and
R. Mart\'inez for technical support



\begin{thebibliography}{}

\bibitem[Abazajian  et al.(2003)]{2003AJ....126.2081A}  Abazajian, K.,
et al. 2003, \aj, 126, 2081

\bibitem[Adelman-McCarthy      et      al.(2006)]{2006ApJS..162...38A}
Adelman-McCarthy, J.~K., et al. 2006, \apjs, 162, 38

\bibitem[Althaus  \&  C{\'o}rsico(2004)]{2004A&A...417.1115A} Althaus,
L.~G., \& C{\'o}rsico, A.~H. 2004, \aap, 417, 1115

\bibitem[Althaus  et al.(2007b)]{2007A&A...467.1175A}  Althaus, L.~G.,
C{\'o}rsico, A.~H., \& Miller Bertolami, M.~M. 2007b, \aap, 467, 1175

\bibitem[Althaus  et  al.(2009)]{2009A&A...494.1021A} Althaus,  L.~G.,
C{\'o}rsico,  A.~H., Torres,  S., \&  Garc{\'{\i}}a-Berro,  E. 2009a,
\aap, 494, 1021

\bibitem[Althaus  et  al.(2003)]{2003A&A...404..593A} Althaus,  L.~G.,
Serenelli,  A.~M.,  C{\'o}rsico, A.~H.,  \&  Montgomery, M.~H.  2003,
\aap, 404, 593

\bibitem[Althaus  et  al.(2008)]{2008ApJ...677L..35A} Althaus,  L.~G.,
C{\'o}rsico, A.~H., Miller  Bertolami, M.~M., Garc{\'{\i}}a-Berro, E.,
\& Kepler, S.~O. 2008, \apjl, 677, L35

\bibitem[Althaus  et al.(2009b)]{2009ApJ...693L..23A}  Althaus, L.~G.,
Garc{\'{\i}}a-Berro, E., C{\'o}rsico,  A.~H., Miller Bertolami, M.~M.,
\& Romero, A.~D. 2009b, \apjl, 693, L23

\bibitem[Althaus  et  al.(2007)]{2007A&A...465..249A} Althaus,  L.~G.,
Garc{\'{\i}}a-Berro, E.,  Isern, J., C{\'o}rsico,  A.~H., \& Rohrmann,
R.~D. 2007a, \aap, 465, 249

\bibitem[Althaus  et  al.(2005)]{2005A&A...435..631A} Althaus,  L.~G.,
Serenelli,     A.~M.,     Panei,     J.~A.,    C{\'o}rsico,     A.~H.,
Garc{\'{\i}}a-Berro, E., \& Sc{\'o}ccola, C.~G. 2005, \aap, 435, 631

\bibitem[Althaus  et  al.(2009)]{2009arXiv0905.1939A} Althaus,  L.~G.,
Panei,  J.~A.,  Romero, A.~D.,  Rohrmann,  R.~D., C{\'o}rsico,  A.~H.,
Garc{\'{\i}}a-Berro, E.,  \& Miller Bertolami, M.~M.  2009c, \aap, 502,
207

\bibitem[Angulo et al.(1999)]{1999NuPhA.656....3A} Angulo, C., et al.
1999, Nuclear Physics A, 656, 3

\bibitem[Benvenuto  \&  Althaus(1999)]{1999ASPC..169..383B} Benvenuto,
O.~G., \& Althaus, L.~G. 1999, MNRAS, 303, 30

\bibitem[Bergeron  et  al.(1995)]{1995PASP..107.1047B}  Bergeron,  P.,
Wesemael, F., \& Beauchamp, A. 1995, \pasp, 107, 1047

\bibitem[Bloecker(1995)]{1995A&A...297..727B}   Bl\"ocker,   T.  1995,
\aap, 297, 727

\bibitem[Burgers(1969)]{1969fecg.book.....B}   Burgers,  J.~M.  1969,
{\sl ``Flow Equations for Composite Gases''}, New York: Academic Press

\bibitem[Cassisi   et  al.(2007)]{2007ApJ...661.1094C}   Cassisi,  S.,
Potekhin, A.~Y., Pietrinferni, A.,  Catelan, M., \& Salaris, M. 2007,
\apj, 661, 1094

\bibitem[Caughlan   \&   Fowler(1988)]{1988ADNDT..40..283C}  Caughlan,
G.~R., \&  Fowler, W.~A. 1988,  Atomic Data and Nuclear  Data Tables,
40, 283

\bibitem[C{\'o}rsico  et  al.(2006)]{2006A&A...458..259C} C{\'o}rsico,
A.~H., Althaus,  L.~G., \& Miller  Bertolami, M.~M. 2006,  \aap, 458,
259

\bibitem[C{\'o}rsico  et  al.(2001)]{2001NewA....6..197C} C{\'o}rsico,
A.~H.,   Benvenuto,    O.~G.,   Althaus,   L.~G.,    Isern,   J.,   \&
Garc{\'{\i}}a-Berro, E. 2001, New Astronomy, 6, 197

\bibitem[Costa \&  Kepler(2008)]{2008A&A...489.1225C} Costa, J.~E.~S.,
\& Kepler, S.~O. 2008, \aap, 489, 1225

\bibitem[Dehner \&  Kawaler(1995)]{1995ApJ...445L.141D} Dehner, B.~T.,
\& Kawaler, S.~D. 1995, \apjl, 445, L141

\bibitem[Diaz-Pinto et al.(1994)]{1994A&A...282...86D} D\'\i az-Pinto,
A., Garc\'\i a--Berro, E., Hernanz, M., Isern, J., \& Mochkovitch, R.
1994, \aap, 282, 86

\bibitem[Dreizler  \& Heber(1998)]{1998A&A...334..618D}  Dreizler, S.,
\& Heber, U. 1998, \aap, 334, 618

\bibitem[Dreizler \&  Werner(1996)]{1996A&A...314..217D} Dreizler, S.,
\& Werner, K. 1996, \aap, 314, 217

\bibitem[Dreizler  et  al.(1997)]{1997fbs..conf..303D}  Dreizler,  S.,
Werner, K.,  Heber, U., Reid,  N., \& Hagen,  H. 1997, in  {\sl ``The
Third  Conference on  Faint  Blue  Stars''}, Eds.:  A.  G. D.  Philip,
J. Liebert, R. Saffer \& D. S.  Hayes (New York: L. Davis Press), 303

\bibitem[Dufour   et   al.(2005)]{2005ApJ...627..404D}   Dufour,   P.,
Bergeron, P., \& Fontaine, G. 2005, \apj, 627, 404

\bibitem[Dufour   et   al.(2007)]{2007Natur.450..522D}   Dufour,   P.,
Liebert, J., Fontaine, G., \& Behara, N. 2007, \nat, 450, 522

\bibitem[Dufour   et   al.(2008)]{2008ApJ...683..978D}   Dufour,   P.,
Fontaine, G., Liebert, J., Schmidt,  G.~D., \& Behara, N. 2008, \apj,
683, 978

\bibitem[Eisenstein  et  al.(2006b)]{2006ApJS..167...40E}  Eisenstein,
D.~J., et al. 2006, \apjs, 167, 40

\bibitem[Fontaine  \&  Brassard(2002)]{2002ApJ...581L..33F}  Fontaine,
G., \& Brassard, P. 2002, \apjl, 581, L33

\bibitem[Fontaine  \&  Brassard(2008)]{2008PASP..120.1043F}  Fontaine,
G., \& Brassard, P. 2008, \pasp, 120, 1043

\bibitem[Fujimoto(1977)]{1977PASJ...29..331F}  Fujimoto,  M.~Y. 1977,
\pasj, 29, 331

\bibitem[Garc\'\i a--Berro et al.(1988)]{1988Natur.333..642G} Garc\'\i
a--Berro, E., Hernanz, M., Isern,  J., \& Mochkovitch, R. 1988, \nat,
333, 642

\bibitem[Gautschy \& Althaus(2002)]{2002A&A...382..141G} Gautschy, A.,
\& Althaus, L.~G. 2002, \aap, 382, 141

\bibitem[Guerrero  et  al.(2004)]{2004A&A...413..257G}  Guerrero,  J.,
Garc{\'{\i}}a-Berro, E., \& Isern, J. 2004, \aap, 413, 257

\bibitem[Haft  et al.(1994)]{1994ApJ...425..222H}  Haft,  M., Raffelt,
G., \& Weiss, A. 1994, \apj, 425, 222

\bibitem[Hansen  et al.(2007)]{2007ApJ...671..380H}  Hansen, B.~M.~S.,
et al. 2007, \apj, 671, 380

\bibitem[Herwig   et   al.(1999)]{1999A&A...349L...5H}   Herwig,   F.,
Bl{\"o}cker, T., Langer, N., \& Driebe, T. 1999, \aap, 349, L5

\bibitem[H{\"u}gelmeyer       et       al.(2005)]{2005A&A...442..309H}
H{\"u}gelmeyer, S.~D.,  Dreizler, S., Werner,  K., Krzesi{\'n}ski, J.,
Nitta, A., \& Kleinman, S.~J. 2005, \aap, 442, 309

\bibitem[H{\"u}gelmeyer       et       al.(2006)]{2006A&A...454..617H}
H{\"u}gelmeyer, S.~D., Dreizler,  S., Homeier, D., Krzesi{\'n}ski, J.,
Werner, K., Nitta, A., \& Kleinman, S.~J. 2006, \aap, 454, 617

\bibitem[]{} Hummer, D. G., \& Mihalas, D. 1988, \apj, 331, 794

\bibitem[Iben et al.(1983)]{1983ApJ...264..605I} Iben, I., Jr., Kaler,
J.~B., Truran, J.~W., \& Renzini, A. 1983, \apj, 264, 605

\bibitem[Iglesias   \&   Rogers(1996)]{1996ApJ...464..943I}  Iglesias,
C.~A., \& Rogers, F.~J. 1996, \apj, 464, 943

\bibitem[Isern    et   al.(2008)]{2008ApJ...682L.109I}    Isern,   J.,
Garc{\'{\i}}a-Berro, E., Torres, S.,  \& Catal{\'a}n, S. 2008, \apjl,
682, L109

\bibitem[Isern et  al.(1998)]{1998ApJ...503..239I} Isern, J., Garc\'\i
a--Berro, E., Hernanz, M., Mochkovitch, R., \& Torres, S. 1998, \apj,
503, 239

\bibitem[Itoh  et al.(1996)]{1996ApJS..102..411I}  Itoh,  N., Hayashi,
H., Nishikawa, A., \& Kohyama, Y. 1996, \apjs, 102, 411

\bibitem[Kawaler   \&   Bradley(1994)]{1994ApJ...427..415K}   Kawaler,
S.~D., \& Bradley, P.~A. 1994, \apj, 427, 415

\bibitem[Kepler  et   al.(2007)]{2007MNRAS.375.1315K}  Kepler,  S.~O.,
Kleinman,   S.~J.,  Nitta,  A.,   Koester,  D.,   Castanheira,  B.~G.,
Giovannini, O.,  Costa, A.~F.~M., \&  Althaus, L. 2007,  \mnras, 375,
1315

\bibitem[Kleinman et  al.(2004)]{2004ApJ...607..426K} Kleinman, S.~J.,
et al. 2004, \apj, 607, 426

\bibitem[Krzesi{\'n}ski       et       al.(2004)]{2004A&A...417.1093K}
Krzesi{\'n}ski,  J.,  Nitta,   A.,  Kleinman,  S.~J.,  Harris,  H.~C.,
Liebert, J., Schmidt,  G., Lamb, D.~Q., \& Brinkmann,  J. 2004, \aap,
417, 1093

\bibitem[Lor{\'e}n-Aguilar      et     al.(2005)]{2005MNRAS.356..627L}
Lor{\'e}n-Aguilar,  P.,  Guerrero,  J.,  Isern, J.,  Lobo,  J.~A.,  \&
Garc{\'{\i}}a-Berro, E. 2005, \mnras, 356, 627

\bibitem[Magni \& Mazzitelli(1979)]{1979A&A....72..134M} Magni, G., \&
Mazzitelli, I. 1979, \aap, 72, 134

\bibitem[Mestel(1952)]{1952MNRAS.112..583M} Mestel,  L. 1952, \mnras,
112, 583

\bibitem[Miller  Bertolami   \&  Althaus  (2006)]{2006A&A...454..845M}
Miller Bertolami, M.~M., \& Althaus, L.~G. 2006, \aap, 454, 845

\bibitem[Miller   Bertolami   \&   Althaus(2007)]{2007A&A...470..675M}
Miller Bertolami, M.~M., \& Althaus, L.~G. 2007, \aap, 470, 675

\bibitem[Miller  Bertolami  et al.(2006)]{2006A&A...449..313M}  Miller
Bertolami, M.~M.,  Althaus, L.~G., Serenelli, A.~M.,  \& Panei, J.~A.
2006, \aap, 449, 313

\bibitem[Miller  Bertolami  et al.(2008)]{2008A&A...491..253M}  Miller
Bertolami,  M.~M., Althaus, L.~G.,  Unglaub, K.,  \& Weiss,  A. 2008,
\aap, 491, 253

\bibitem[O'Brien   \&   Kawaler(2000)]{2000ApJ...539..372O}   O'Brien,
M.~S., \& Kawaler, S.~D. 2000, \apj, 539, 372

\bibitem[Prada  Moroni \&  Straniero(2007)]{2007A&A...466.1043P} Prada
Moroni, P.~G., \& Straniero, O. 2007, \aap, 466, 1043

\bibitem[Rauch  et al.(2006)]{2006ASPC..348..194R}  Rauch,  T., Reiff,
E., Werner,  K., Herwig, F., Koesterke,  L., \& Kruk,  J.~W. 2006, in
{\sl ``Astrophysics  in the Far  Ultraviolet: Five Years  of Discovery
with  FUSE''}, Eds.: G.  Sonneborn, H.  Moos \&  B. G.  Andersson, ASP
Conf. Ser., 348, 194

\bibitem[Rohrmann et  al.(2002)]{2002MNRAS.335..499R} Rohrmann, R.~D.,
Serenelli, A.~M.,  Althaus, L.~G., \& Benvenuto,  O.~G. 2002, \mnras,
335, 499

\bibitem[Saio  \&  Jeffery(2000)]{2000MNRAS.313..671S}  Saio,  H.,  \&
Jeffery, C.~S. 2000, \mnras, 313, 671

\bibitem[Sch\"onberner(1979)]{1979A&A....79..108S}  Sch\"onberner, D.
1979, \aap, 79, 108

\bibitem[Segretain  et al.(1994)]{1994ApJ...434..641S}  Segretain, L.,
Chabrier,   G.,  Hernanz,   M.,  Garcia-Berro,   E.,  Isern,   J.,  \&
Mochkovitch, R. 1994, \apj, 434, 641

\bibitem[Straniero  et al.(2003)]{2003ApJ...583..878S}  Straniero, O.,
Dom{\'{\i}}nguez, I., Imbriani, G., \& Piersanti, L. 2003, \apj, 583,
878

\bibitem[Tassoul   et  al.(1990)]{1990ApJS...72..335T}   Tassoul,  M.,
Fontaine, G., \& Winget, D.~E. 1990, \apjs, 72, 335

\bibitem[Torres   et   al.(2002)]{2002MNRAS.336..971T}   Torres,   S.,
Garc{\'{\i}}a--Berro,  E., Burkert,  A., \&  Isern, J.  2002, \mnras,
336, 971

\bibitem[Unglaub  \& Bues(1998)]{1998A&A...338...75U} Unglaub,  K., \&
Bues, I. 1998, \aap, 338, 75

\bibitem[Unglaub  \& Bues(2000)]{2000A&A...359.1042U} Unglaub,  K., \&
Bues, I. 2000, \aap, 359, 1042

\bibitem[van   Hoof  et   al.(2007)]{2007A&A...471L...9V}   van  Hoof,
P.~A.~M., et al. 2007, \aap, 471, L9

\bibitem[van  Horn(1968)]{1968ApJ...151..227V} van Horn,  H.~M. 1968,
\apj, 151, 227

\bibitem[Werner(2001)]{2001Ap&SS.275...27W}  Werner, K.  2001, \apss,
275, 27

\bibitem[Werner  \& Herwig(2006)]{2006PASP..118..183W} Werner,  K., \&
Herwig, F. 2006, \pasp, 118, 183

\bibitem[Werner et  al.(2008)]{2008ASPC..391..239W} Werner, K., Rauch,
T.,  \& Kruk, J.~W.  2008a, in  {\sl ``Hydrogen-Deficient  Stars''} ,
Eds.: K. Werner \& T. Rauch, ASP Conf. Ser., 391, 239

\bibitem[Werner et  al.(2008)]{2008A&A...492L..43W} Werner, K., Rauch,
T., \& Kruk, J.~W. 2008b, \aap, 492, L43

\bibitem[Werner et  al.(2004)]{2004A&A...424..657W} Werner, K., Rauch,
T., Napiwotzki, R., Christlieb, N., Reimers, D., \& Karl, C.~A. 2004,
\aap, 424, 657

\bibitem[Winget  \& Kepler(2008)]{2008ARA&A..46..157W}  Winget, D.~E.,
\& Kepler, S.~O. 2008, \araa, 46, 157

\bibitem[Winget  et   al.(2009)]{2009ApJ...693L...6W}  Winget,  D.~E.,
Kepler, S.~O.,  Campos, F., Montgomery, M.~H.,  Girardi, L., Bergeron,
P., \& Williams, K. 2009, \apjl, 693, L6

\bibitem[Winget  et   al.(1987)]{1987ApJ...315L..77W}  Winget,  D.~E.,
Hansen,  C.~J., Liebert, J.,  van Horn,  H.~M., Fontaine,  G., Nather,
R.~E., Kepler, S.~O., \& Lamb, D.~Q. 1987, \apjl, 315, L77

\bibitem[Wood(1995)]{1995whdw.conf...41W}  Wood, M.~A.  1995, Lecture
Notes in Physics, (Berlin: Springer-Verlag) Vol.~443, 41

\bibitem[York et al.(2000)]{2000AJ....120.1579Y}  York, D.~G. et al.
2000, \aj, 120, 1579

\end{thebibliography}
\end{document}